\documentclass{article}
\usepackage{graphicx} % Required for inserting images
\usepackage{amssymb}
\usepackage{pifont}
\usepackage{enumitem}
\usepackage{tikz}
\usepackage{xcolor}
\usepackage{geometry}

\usepackage{amsmath} 
\usepackage{hyperref}
\usepackage{float}
\usepackage{tabularx} 
\usepackage{gensymb}
\usepackage{textcomp}
\usepackage{caption}
\usepackage{subcaption}
\usepackage{array}
\usepackage{multirow}
\usepackage{listings}
\usepackage[dvipsnames]{xcolor}
\usepackage{pythonhighlight}
\usepackage{tocloft}
\usepackage{fancyhdr}
\usepackage{mathtools}
\usepackage{pgfplots}
\usepackage{changepage}
\usepackage{pgfplotstable}
\usepackage{tipa}
\usepackage{setspace}
\usepackage{soul}
\usepackage{capt-of}
\usepackage{afterpage}
\usepackage{xcolor}
\usepackage[percent]{overpic}
\usepackage{authblk}
\usepackage{etoolbox}
\makeatletter
\patchcmd{\@bibitem}
  {\newblock}
  {\newblock\bfseries}
  {}{}
\makeatother

\usepackage[super]{natbib} 
\bibpunct{}{}{,}{s}{}{}

\graphicspath{{Figures/}}

\pgfplotsset{compat=1.18}
\fontsize{12pt}{12pt}\selectfont

\begin{document}
\title{A complex network approach to characterize clustering of events in irregular time series}

\author[1,2]{Ambedkar Sanket Sukdeo}
\author[1,2]{K. Shri Vignesh\thanks{kshrivignesh1995@gmail.com}}
\author[3]{Sachin S. Gunthe}
\author[4]{T Narayan Rao}
\author[4]{Amit Kumar Patra}
\author[1,2]{R. I. Sujith}

\affil[1]{Department of Aerospace Engineering, Indian Institute of Technology Madras}
\affil[2]{Center of Excellence for studying Critical Transitions in Complex Systems}
\affil[3]{Environmental Engineering Division, Department of Civil Engineering, Indian Institute of Technology Madras, Chennai, India}
\affil[4]{National Atmospheric Research Laboratory, Gadanki, India}

\date{}
\maketitle

\section*{Abstract}

\large In complex systems, events occur at irregular intervals that inherently encode the underlying dynamics of the system. Analyzing the temporal clustering of these events reveals critical insights into the non-random patterns and the temporal evolution. Existing techniques can effectively quantify the overall clustering tendency of events using global statistical measures. However, these macroscopic approaches leave a critical gap, as they do not attempt to investigate the dynamics of individual clusters. Analyzing individual clusters is essential, as it helps comprehend the local interactions that actively drive the system dynamics, which may be obscured by global averaging, while simultaneously revealing the time scales involved. To address these limitations, we propose a complex network-based framework for analyzing clustering of events occurring at irregular intervals. The framework establishes connections using arrival times, transforming the time series into a network. Network properties are then used to quantify the clustering. Further, a community detection algorithm is used to identify individual clusters in time series. We illustrate the method by applying it to standard arrival processes, such as the Poisson process and the Markov-modulated Poisson process. To further demonstrate its scope, we apply the method to two diverse systems: the time series of droplet arrivals in turbulent flows and the R-R intervals in electrocardiogram (ECG) signals.\\

\section{Introduction} \label{intro}
\large
Arrival processes characterize the way discrete events unfold in time, capturing the unique timing associated with a sequence of individual occurrences. Understanding these arrival patterns is essential, as they directly reflect the underlying dynamics of the system being studied.\cite{daw2018queues} They occur across a wide range of systems, where the timing of occurrence of events carries critical information, for example, in nature the occurrence of natural hazards like earthquakes \cite{ogata1988statistical}, lightning, and volcanic eruptions, trading transactions in financial markets \cite{reno2003closer,azizpour2018exploring,zhang2004model}, user clicks in webpage  \cite{SU20151,singh2021clustering}, occurrences of critical signals by electronic healthcare equipment \cite{geva2017brain}, such as ECG. Understanding the timing of occurrences of these events allows us to explore the underlying patterns and system dynamics that are essential for tasks such as probabilistic hazard forecasting and personalized recommendations. These arrival processes may occur at fixed, regular intervals or at irregular, varying intervals.
\\
\large Processes with randomly occurring events are typically modeled using Poisson processes \cite{prekopa1957poisson}, which assume that events are memoryless, occur independently, and follow a constant average rate of occurrence over time. However, empirical evidence suggests that real-world phenomena rarely adhere to this pure randomness. Instead, the arrivals are heavily influenced by the past occurrences and memory effects, That manifest later in future events. In such irregular time series, events often occur in clusters of closely spaced arrivals, followed by extended periods of inactivity. This clustering phenomenon is observed across diverse systems-for example, seismic aftershocks closely follow major earthquakes \cite{ogata1988statistical,telesca2007time}, rapid sell-offs cascade through stock markets \cite{zhang2004model}, and viral trends spread explosively across social media platforms.\cite{wuebben2016getting} Here, quantifying clustering among events helps reveal the underlying mechanisms that govern the system processes.
\\
\large Traditionally, clustering among events in time series have been quantified using global statistical measures. The Fano factor and Allan factor are commonly used to evaluate clustering in arrival processes by measuring deviations from Poisson behavior.\cite{telesca2007time} Both methods exploit the fundamental property of the Poisson process, namely the equality between the mean and variance of the number of arrivals within a given temporal window. Telesca \cite{telesca2007time} used the Fano factor and Allan factor to characterize event clustering in natural hazards, quantifying the extent to which events tend to cluster together in time. Another tool used by researchers to quantify clustering in time series is fishing statistic.\cite{baker2010analysis} Similar to the Fano and Allan factors, the fishing statistic evaluates deviations from a Poisson process using the dispersion index, defined as the ratio of the variance of event counts within a specific bin to their mean. For a Poisson process, the expected value of the dispersion index is unity. The fishing statistic functions as a hypothesis test by subtracting this expected Poisson value from the dispersion index and normalizing the result by the theoretical standard deviation of the dispersion index. Baker \textit{et al.} \cite{baker2010analysis} utilized the fishing statistic to detect and quantify the spatial clustering of cloud droplets in turbulent flow from the droplet arrival time series acquired from an aircraft.
\\
\large Real-world systems exhibit inherent complexity, with multiple subsystems interacting dynamically across a wide range of spatial and temporal scales. Clouds, for instance, embody such multiscale interactions, where different phases of water couple with turbulent flows.\cite{baker1992turbulent,uhlig1998holographic} Characterizing microscale droplet clustering is particularly crucial, as the spatiotemporal organization of droplets directly influences macroscopic cloud properties, including droplet collisions, coalescence, and precipitation. Given these complexities, a single global clustering measure cannot adequately capture the system’s dynamics, and the relevant time scales must also be resolved. It is therefore highly desirable to employ clustering measures at both the event scale and the global scale, while simultaneously identifying distinct highly clustered regions and assigning a dedicated measure to each. Accurately identifying and quantifying these clusters can drastically improve our understanding and predictive capabilities regarding the true behavior of a system. Complex networks provide a powerful framework for analyzing system dynamics across multiple scales. In this representation, the individual components of a system are treated as nodes, with links established between them based on their interactions. This approach enables analysis from the smallest scales to the largest scales: small-scale interactions at the node level can collectively give rise to emergent large-scale dynamics across the entire system. This ability to link local interactions to overall system behavior has made complex networks a widely used approach in studying various complex systems, such as climate dynamics \cite{marwan2015complex}, biological processes \cite{bassett2017network}, and fluid mechanics.\cite{shri2022complex, tandon2023multilayer}
\\
\large We propose a complex network approach to analyze the clustering of events in irregular time series by mapping the time series onto a network. Each occurrence of an event is represented as a node, and edges are established based on interactions between nodes. This framework enables a multiscale analysis that spans from individual events and localized clusters to comprehensive global metrics.  We use community detection algorithms to identify groups of cluster of occurrences. Thus, by bridging different levels, the complex network approach offers a comprehensive analysis into the structure of the system, dynamics, emergent properties, and patterns. Using several network measures, we characterize the individual clusters and quantify the strength of clustering and unravel the time scales associated with them.
\\
\large The subsequent sections of the paper are organized as follows: Section \ref{Sec: Methodology} provides a detailed description of the network construction from irregular time series. The event clustering analysis of standard arrival processes is described in Sec. \ref{Sec: standard arrival processes}. The analysis of the time series of droplet arrivals to understand clustering is discussed in Sec. \ref{Sec: PDPA time series}, and the application of the method to detect heart arrhythmia is illustrated in Sec. \ref{Sec: Heart}. Finally, we discuss and conclude in Sec. \ref{Sec: Conclusions}.

\graphicspath{{figures/}}
\section{Methods} \label{Sec: Methodology}

\subsection{Construction of complex network}
\label{Sec: Method_Construction of complex network}

\large A complex network consists of nodes that are connected with each other using links. The links are established by defining a relation between the nodes. We construct a complex network from an irregular time series by considering each arrival in the time series as a node in the network. The time series consists of arrival times for every arrival. For example, arrival $i$ happens at time $T_{\mathrm{i}}$ and arrival $j$ happens at time $T_{\mathrm{j}}$. The interarrival time is the time between two arrivals. The interarrival time $(t_{ij})$ for arrivals $i$ and $j$ is given by,
\begin{equation}\label{Eq: Interarrival time}
    \begin{aligned}
        t_{ij} = \left| T_{i} - T_{j} \right|,
    \end{aligned}
\end{equation}
The average arrival rate $(AR)$ for an irregular time series is defined as the ratio of total number of arrivals $(n)$ to total time $(T_{n})$; i.e.,
\begin{equation}\label{Eq: Average arrival rate}
    \begin{aligned}
        AR=\frac{n}{T_{n}}.
    \end{aligned}
\end{equation}

\begin{figure*}
    \centering
    \includegraphics[width=1\linewidth]{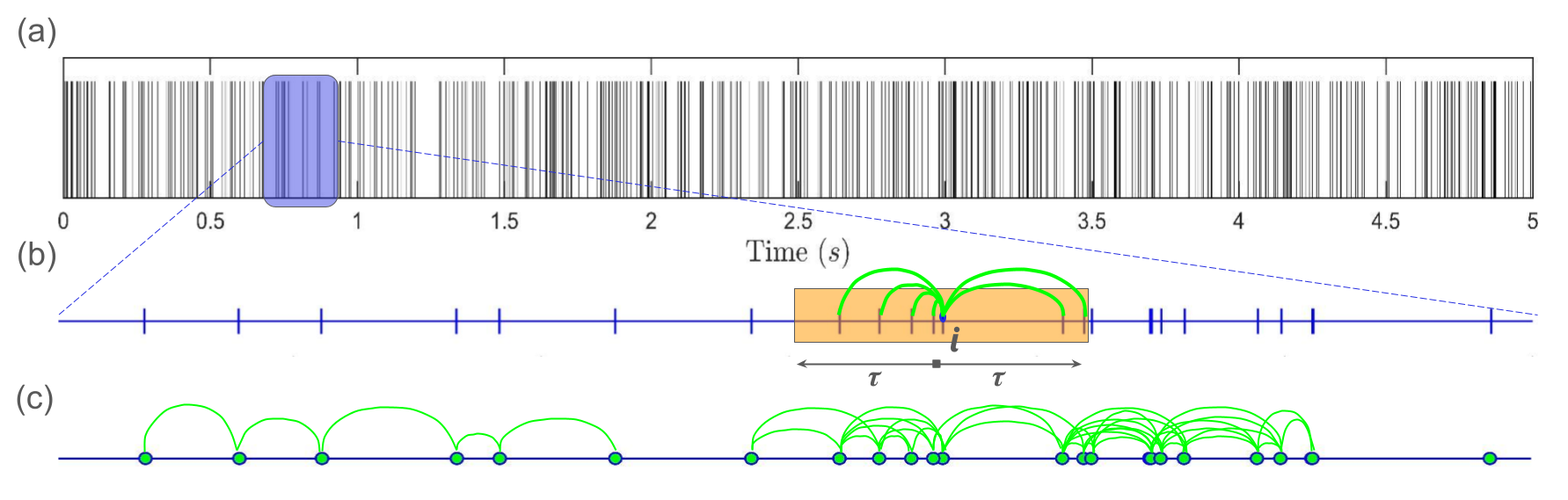}
    \caption{Schematic of construction of network from irregular arrival time series. \textbf{(a)} Irregular arrival time series with arrivals represented as vertical lines.  \textbf{(b)} Section of the time series showing network construction for arrival $i$ with time window $\tau$ forward and backward.  \textbf{(c)} Section of the time series represented as nodes and links after completing the network construction.} 
    \label{Fig: Const of complex network}
\end{figure*}

\large Figure~\ref{Fig: Const of complex network} illustrates the network construction from an irregular time series. Let us consider node $N_{i}$ corresponding to arrival $i$ (see Fig~\ref{Fig: Const of complex network} (b)) happening at time $T_{i}$. We connect all arrivals within the time interval $\tau$ to node $N_{i}$, where $\tau$ is the inverse of the average arrival rate. Note that we evaluate time $\tau$ in both forward and backward directions from the considered arrival $i$. We assign each link a weight equal to the inverse of the interarrival time. Thus, we have a weighted and undirected network. The links and weights of the links are then represented using an adjacency matrix, where each matrix element denotes the weight of the link between a pair of nodes. Each element of the adjacency matrix is calculated using the following equation,
\begin{equation}\label{Eq: Adjacency matrix}
    \begin{aligned}
        A_{ij} = \left\{ \begin{array}{ll}
                    \frac{1}{t_{ij}}, &  t_{ij} < \tau \\
                     0, & otherwise.
                \end{array}\right.
    \end{aligned}
\end{equation}
\\

\subsection{Node strength - a measure of clustering}
\label{Sec: Average NS}

\large The node strength $(S)$ for each node in the network is calculated by adding all the weights for the node. The node strength is normalized by the average arrival rate of the time series. Node strength quantifies how clustered the events are during the occurrence of the corresponding node. The average node strength $(S_{avg})$ for a time series is the mean of the node strengths of all nodes in the network. The calculation of the node strength and the average node strength is given by,
\begin{equation}\label{Eq: Node Strength}
    \begin{aligned}
        S_{i} = \frac{\sum_{j=1}^{n} A_{ij}}{AR},
    \end{aligned}
\end{equation}
\begin{equation}\label{Eq: Avg Node Strength}
    \begin{aligned}
        S_{avg} = \frac{\sum_{i=1}^{n} S_{i}}{n}.
    \end{aligned}
\end{equation}
The average node strength gives a global estimate of the clustering of events in the time series, while the node strength of a particular node provides a local estimate of clustering during the occurrence of an event. Equation~\ref{Eq: Flow chart} shows the flow chart for the calculation of node strength and the average node strength from the adjacency matrix obtained after network construction.
\begin{equation}\label{Eq: Flow chart}
    \begin{aligned}
        A_{ij}=\begin{bmatrix}
A_{11} & A_{12} & \cdots & A_{1n} \\
A_{21} & A_{22} & \cdots & A_{2n} \\
\vdots & \vdots & \ddots & \vdots \\
A_{n1} & A_{n2} & \cdots & A_{nn}\end{bmatrix}
\longrightarrow S_{i}= \begin{bmatrix} S_{1} \\ S_{2} \\ \vdots\\S_{n} \end{bmatrix} \longrightarrow S_{avg}
    \end{aligned}
\end{equation}
\\

 \subsection{Community detection - locating individual clusters}
\label{Sec: Community detection}

\large We use community detection to identify groups of highly clustered events in the time series. Highly interconnected nodes are identified based on their structural characteristics from the constructed network and are marked as individual clusters of events.\cite{yang2013overlapping,kelley2012defining} Several methods for community detection have been developed, with applications ranging across various fields, such as biology, physics, social sciences, applied mathematics, and computer science.\cite{lancichinetti2009community} Among these, modularity-based methods are widely used.\cite{newman2004fast,newman2004finding} These methods try to maximize the modularity of the networkwhile organizing nodes into communities.\cite{blondel2008fast} Modularity $(Q)$ is given as,
\begin{equation}\label{Eq: Modularity}
    \begin{aligned}
        Q = \frac{1}{2m}\sum_{ij}^{} \left[ A_{ij}-\frac{S_{i}S_{j}}{2m} \right]\delta(c_{i},c_{j})  ,
    \end{aligned}
\end{equation}
where, $A_{ij}$ is the weight of edge between $i$ and $j$, $S_{i}$ is the strength of node $i$, $c_{i}$ is the community to which node $i$ is assigned, the $\delta$ function $\delta(u, v)$ is 1 if $u = v$ and 0 otherwise and $m=\frac{1}{2} \sum_{ij} A_{ij}$. 
\\
\large We use the algorithm introduced by Blondel \textit{et al.} \cite{blondel2008fast} (Louvain's algorithm) to identify communities in our complex network. The algorithm starts with assigning a separate community to each node. Hence, at the beginning, the number of communities is the same as the number of nodes in the network. Then, for each node $i$, we calculate the gain of modularity for removing $i$ from its community and placing it in the community of neighbors $j$. The node $i$ is then placed in the community where the gain in modularity is positive and maximum. The node $i$ stays in its original community if no positive gain is possible. This process is repeated for all nodes sequentially until no further gain in modularity can be achieved. One advantage of this method is that it does not require prior knowledge of the number of communities. The communities detected by this algorithm are individual clusters in irregular time series. Upon identifying individual clusters, their average node strength and associated time scales can be examined to better understand system dynamics across different clusters. To summarize, the constructed network enables the use of network measures at different levels to investigate clustering among events in the time series across multiple scales.
\section{Clustering of events in standard arrival processes} 
\label{Sec: standard arrival processes}

\begin{figure}[h]
    \centering
    \includegraphics[width=0.75\linewidth]{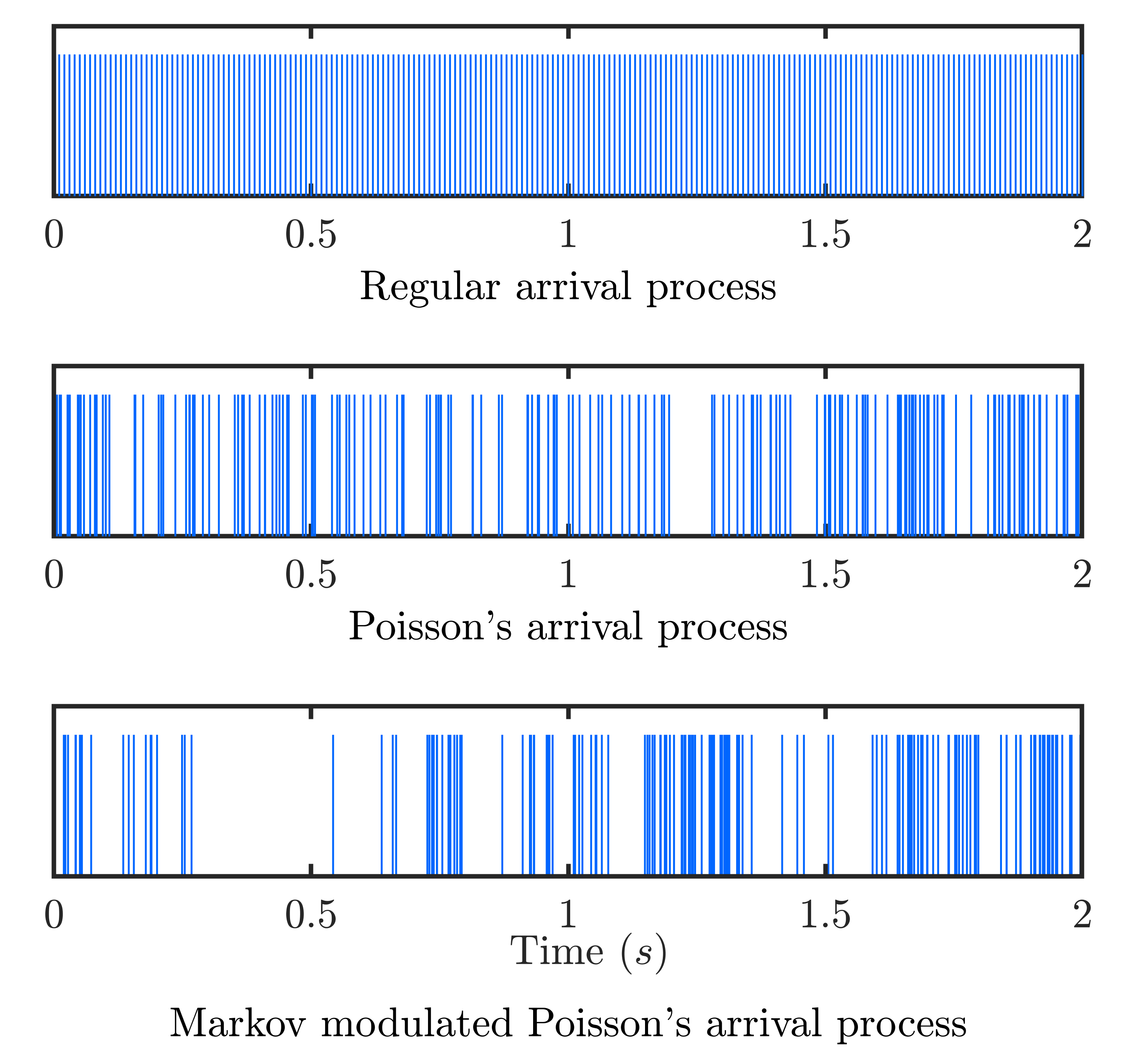}
    \caption{Time series generated for different arrival processes. The vertical lines represent arrivals in a time series. A two-second section with 200 arrivals in each time series is shown. Arrivals are equally spaced for the regular arrival process. The periods of inactivity and arrivals are much closer in MMPP, than in Poisson's arrival process.}
    \label{Fig: All Time series}
\end{figure}

\large To illustrate and validate our approach, we analyzed standard arrival processes such as regular arrivals, Poisson's arrival process\cite{prekopa1957poisson}, and Markov-modulated Poisson process (MMPP).\cite{cinlar1975introduction} The regular arrival process has a constant interarrival time between each consecutive arrival. Prékopa \cite{prekopa1957poisson} demonstrated that an arrival process with independent increments, under suitable conditions, exhibits properties consistent with a Poisson process. In Poisson's arrival process, the interarrival times follow an exponential distribution.\cite{daw2018queues} MMPP is a doubly stochastic Poisson's process in which interarrival times follow the Poisson's arrival process while the arrival rate changes according to the Markov chain.\cite{fischer1993markov} These arrival processes have been widely studied in queuing theory and stochastic modeling\cite{benes2017general,ibe2013markov}, making them suitable for testing our method. For each process, we generated and analyzed 100 independent realizations of the time series, each containing 10,000 arrivals, with a fixed mean arrival rate. Appendix \ref{Apend: A} explains the procedure for generating the time series for each arrival process. A section of the generated time series is shown in Fig.~\ref{Fig: All Time series}. We construct complex networks from each time series and analyzed different network measures.

\begin{figure}
    \centering
    \includegraphics[width=0.75\linewidth]{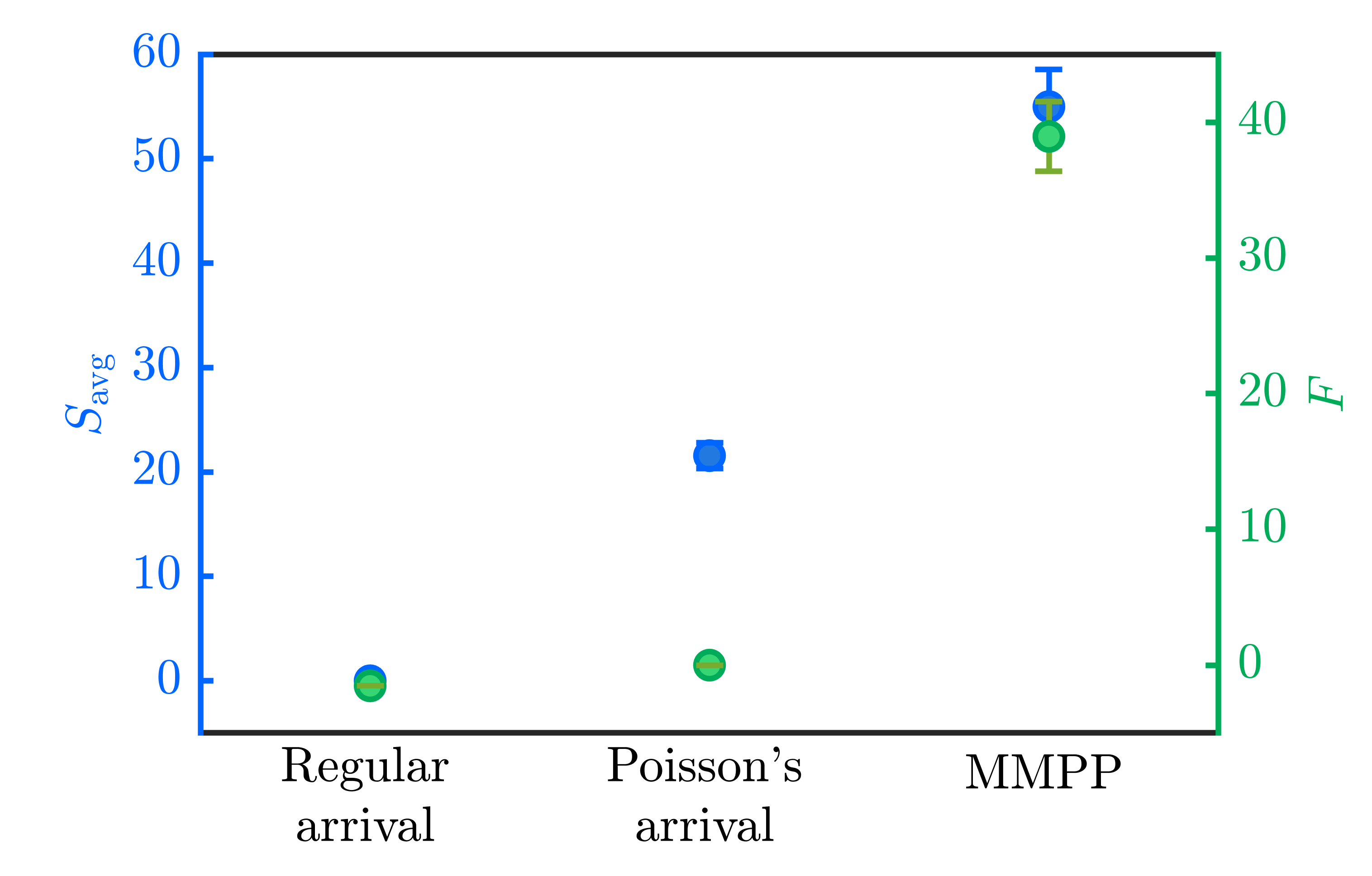}
    \caption{Comparison of global clustering measures average node strength $S_{avg}$ and fishing statistic $F$. Both measures are zero for regular intervals. For Poisson arrival process, $F$ is zero because it measures deviations from the Poisson arrival processes, while the $S_{avg}$ provides the absolute value of clustering. For MMPP, both measures accurately estimate the higher levels of clustering observed.}
    \label{Fig: FTvsSavg}
\end{figure}

\large We can clearly observe that the arrival time series following MMPP arrivals has the maximum clustering. The random nature of arrivals in a Poisson process produces an apparent degree of clustering among events, this clustering remains statistically constant across all Poisson processes, in contrast to the more uniform spacing observed in the regular arrival processes. We compare the global clustering measure ($S_{avg}$) in a time series with the fishing statistic \cite{baker1992turbulent} to validate our approach. A comparison of the average node strength $(S_{avg})$ and the fishing statistic $(F)$ for each arrival process is plotted in Fig.~\ref{Fig: FTvsSavg}. As expected, we see that both the $S_{avg}$ and $F$ are nearly zero for regular arrivals. For the Poisson arrival process, the fishing statistic yields a value of zero, since it is a hypothesis-testing measure designed to quantify deviations from the Poisson arrival process.\cite{baker2010analysis} Meanwhile, the $S_{avg}$ provides the absolute value of clustering observed in a Poisson arrival process. Note that this value can be used to normalize and quantify the deviations from a Poisson process. The enhanced clustering observed in the MMPP time series is well captured by both the global clustering measures $S_{avg}$ and $F$. Thus, $S_{avg}$ serves as an effective estimate of global clustering among events in a time series.

\begin{figure}
    \centering
    \includegraphics[width=0.75\linewidth]{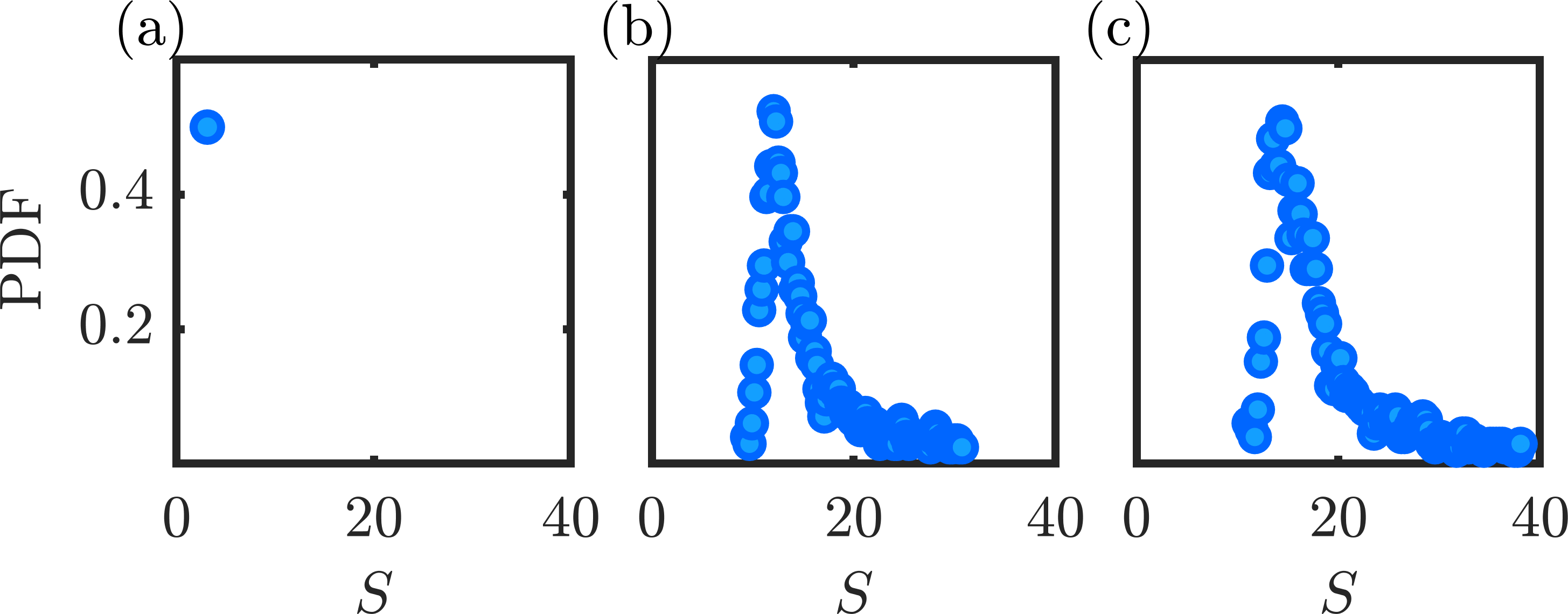}
    \caption{Node strength distribution for \textbf{(a)} regular arrival process, \textbf{(b)} Poisson’s arrival process, and \textbf{(c)} MMPP. In the case of regular arrivals, all nodes have the same node strength. For the Poisson arrival process and MMPP, the distribution is not concentrated at a single value, indicating variability in node strengths, with MMPP showing comparatively higher values.}
    \label{Fig: Node strength}
\end{figure}

\large The node strength of an event quantifies the local clustering among the events during its occurrence. Figure~\ref{Fig: Node strength} shows the node strength distribution for each arrival process. As expected, the node strength distribution for regular arrivals is a discrete point (Fig.~\ref{Fig: Node strength} (a)), signifying that all nodes have the same node strength. For Poisson’s process, most of the nodes have similar node strength as evident from the narrow distribution (Fig.~\ref{Fig: Node strength} (b)), with very few nodes having higher values. Since the MMPP is constructed by varying the arrival rate of a Poisson process, the resulting node strength distribution closely resembles that of the Poisson process, with a shift towards higher node strength (Fig.~\ref{Fig: Node strength} (c)), due to the enhanced clustering in the time series.

\begin{figure}
    \centering
    \includegraphics[width=1\linewidth]{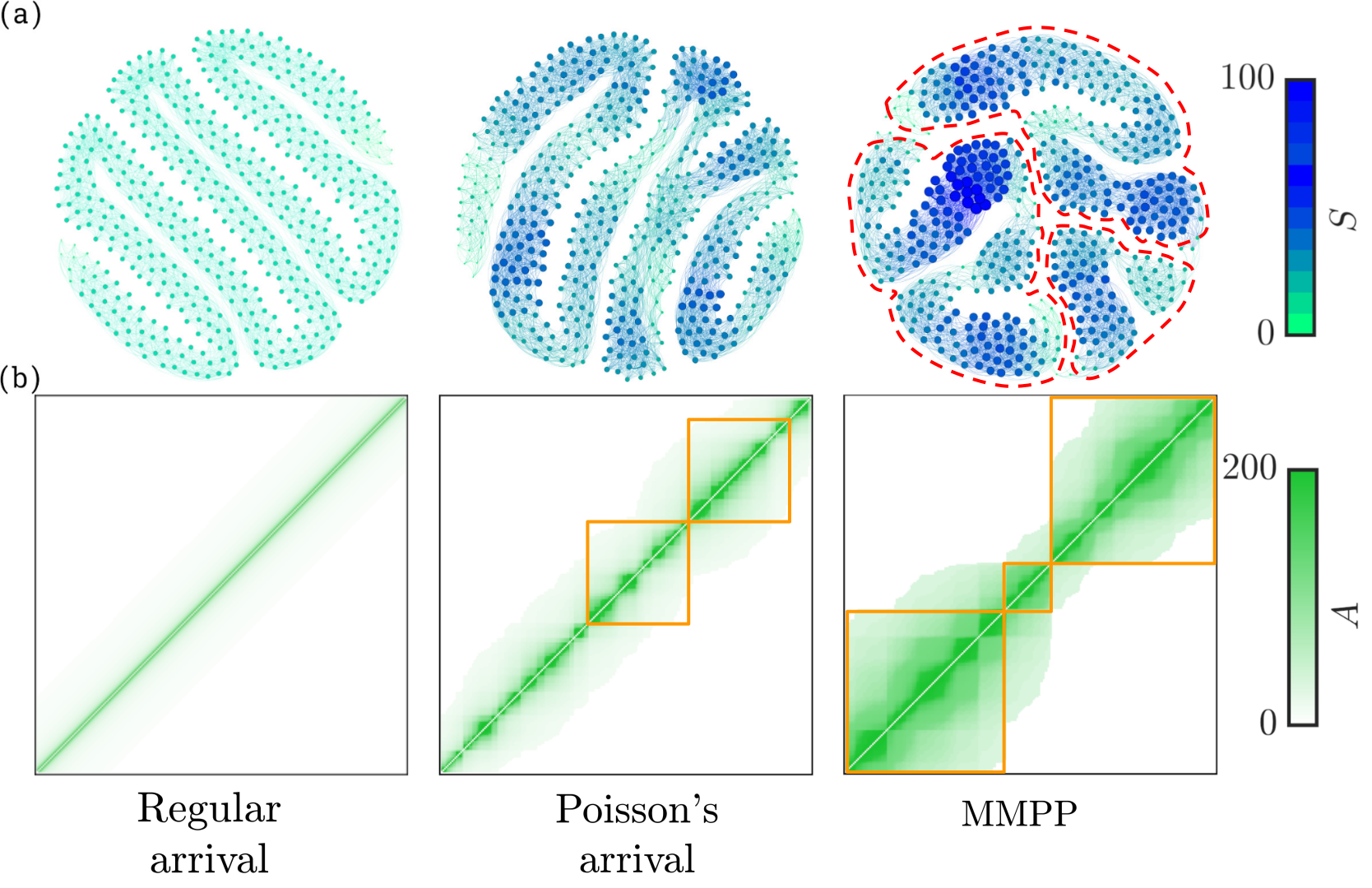}
    \caption{The network structure and the adjacency matrix for each arrival process. \textbf{(a)}  The networks are drawn using Fruchterman and Reingold algorithm\cite{fruchterman1991graph} in Gephi visualization software.\cite{bastian2009gephi} Visualization of the complex networks show that MMPP has locations of packed node structures (nodes with dark blue color), and the network is divided into separate groups (groups are separated by red dashed borders). The Poisson's arrivals have a few packed structures; however, there are no separate groups. The regular arrival process also has no groups and is seen as a connected network. \textbf{(b)} The adjacency matrix with detected communities marked by the orange boxes is plotted for each process. We observe no communities for regular arrivals. For Poisson’s arrival process, there is an emergence of a few small communities, whereas larger communities are seen for MMPP.}
    \label{Fig: CN and ADJ community}
\end{figure}

\large The visualization of complex networks can help us understand the structure and give more insights. We use Gephi software\cite{bastian2009gephi} for visualization. The visualization of the complex network and adjacency matrix for each arrival process is shown in Fig.~\ref{Fig: CN and ADJ community}. We observe in Fig.~\ref{Fig: CN and ADJ community}(a)  that the network of the regular arrival process is well-connected and forms a continuous pattern. In the network for Poisson's process, we observe a few locations with nodes having high node strength while the network is still continuous without any breaks. In contrast, for MMPP, we observe that the network is divided into different groups, with higher frequency of greater node strength values.

\begin{figure}[h]
\centering
\includegraphics[width=0.75\linewidth]{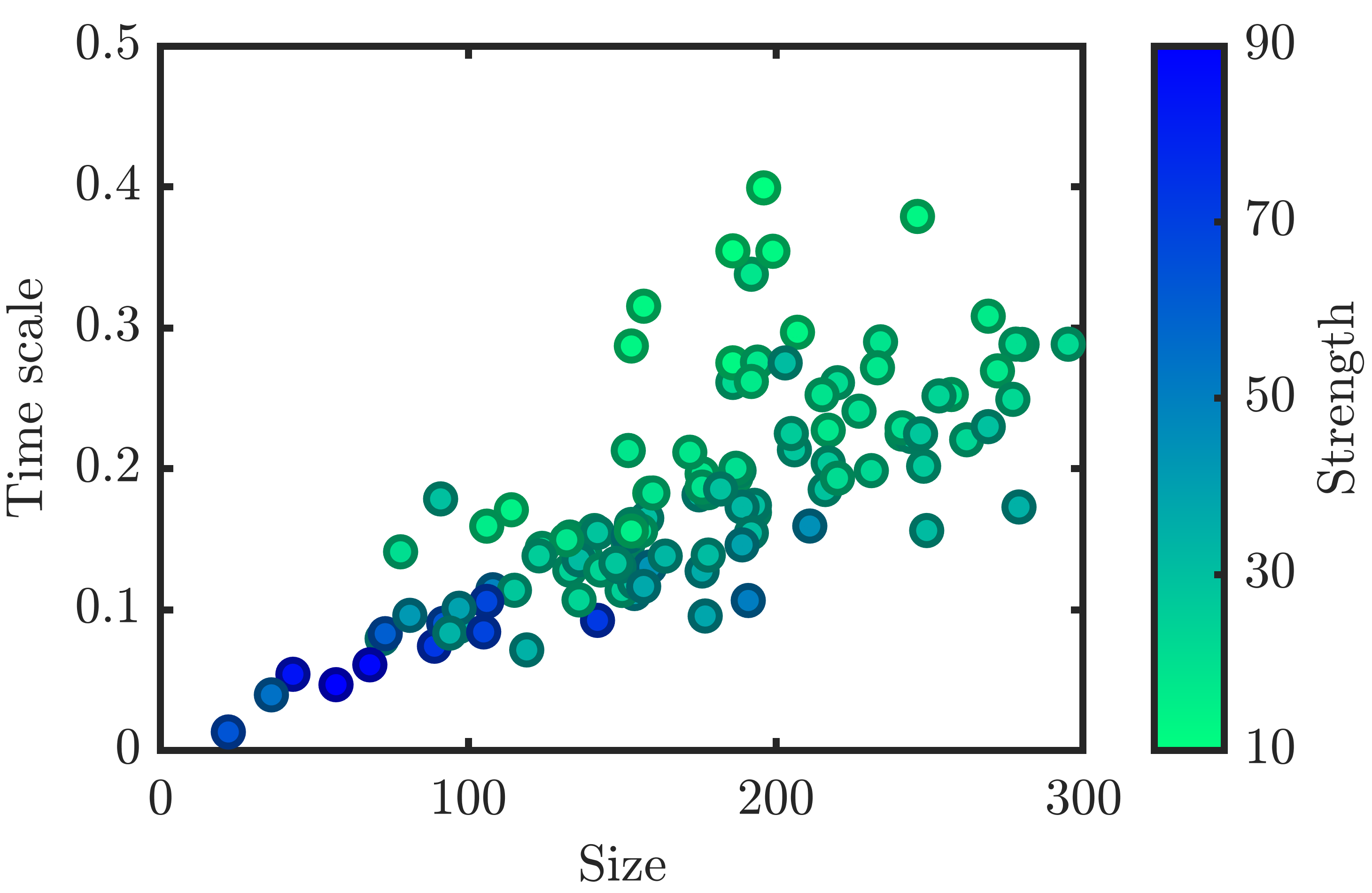}
\caption{Time scales and size of identified clusters for MMPP time series with strength indicated as the color scale. Clusters with smaller size occurring in shorter time scales have higher cluster strength.}
    \label{Fig: Ind. Cluster}
\end{figure}

\large Individual clusters of arrival events are identified employing the modularity maximization algorithm in the complex network. These clusters are group of nodes that have stronger connections within the group than with the rest of the network. Figure ~\ref{Fig: CN and ADJ community}(b) shows the adjacency matrix along with the identified communities marked by the orange boxes. We do not observe any communities for regular arrivals, signifying that there is no cluster of arrivals in the case of regular arrivals. We see a few community structures emerge for Poisson’s arrival, whereas for MMPP, the communities are more significant and bigger than Poisson’s arrival process.
\\
\large Upon identifying the individual clusters in the MMPP time series, we analyze their characteristics, by studying their time scales, size and cluster strength within each cluster. Here, size is the number of arrivals in a given cluster, the time scale of a cluster is the time between the first and the last arrival for the given cluster, and cluster strength is the average of node strengths from all the nodes present in the cluster. The time scale of a cluster unravels how long or short-lived the cluster is, giving insight into the life span of clusters. Figure \ref{Fig: Ind. Cluster} shows the scatter plot of the time scale and size of the cluster with the color indicating the cluster strength. For an MMPP time series, the clusters with smaller size (i.e., fewer arrivals) and shorter time scales have higher cluster strength than those with larger size (i.e., many arrivals) and longer time scales.

\begin{figure}[h]
    \centering
    \includegraphics[width=0.75\linewidth]{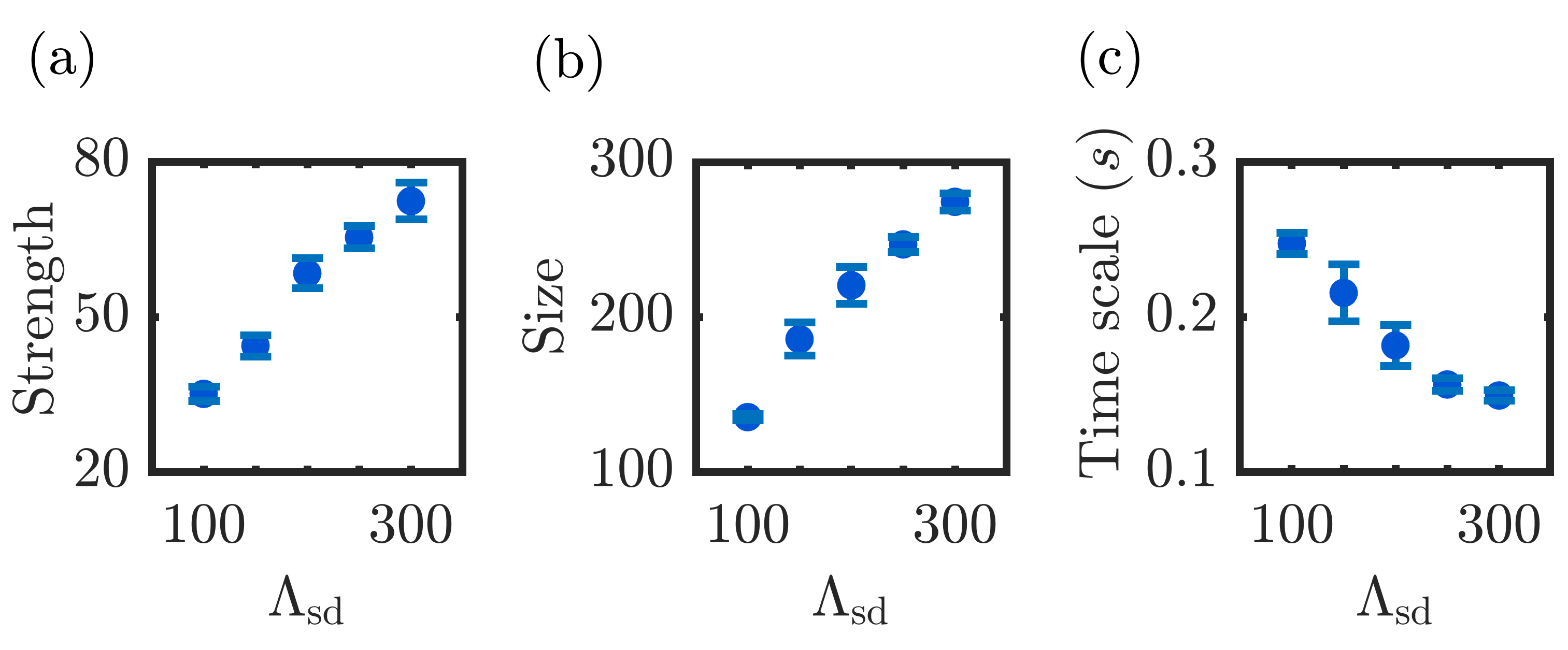}
    \caption{Variation of cluster \textbf{(a)} strength, \textbf{(b)} size and \textbf{(c)} time scale with change in $\Lambda_{sd}$. The cluster strength and size of clusters increase with an increase in $\Lambda_{sd}$, whereas time scales decrease with an increase in  $\Lambda_{sd}$. This shows that short-lived, big, and intense clusters are formed with an increase in $\Lambda_{sd}$.}
    \label{Fig: Variation in Ind. Cluster}
\end{figure}

\large Further, to understand effect of MMPP parameters on clustering, we study five different cases of MMPP time series by varying parameters. The MMPP parameters and generation of different time series are explained in Appendix \ref{Apend: A}. We keep the mean of Poisson’s arrival rate array ($\Lambda_{m}$) constant and only change the standard deviation ($\Lambda_{sd}$) from 100 to 300 in steps of 50. Increasing $\Lambda_{sd}$ increases the variability of the Poisson's arrival rates in MMPP, giving rise to more clustering. We construct the complex network, apply community detection, and compute the cluster size, the cluster strength, and the time scale of clusters for each case of the MMPP time series. This procedure is repeated for 100 surrogate time series in each case, and the variation of the average cluster size, strength, and time scale are shown in Fig.~\ref{Fig: Variation in Ind. Cluster}. We observe in Fig. \ref{Fig: Variation in Ind. Cluster} that the strength and size of clusters increase with an increase in $\Lambda_{sd}$, whereas the time scale of clusters decreases. This implies that for high $\Lambda_{sd}$, there are many closeby arrivals in a very short time span, giving rise to stronger clustering. The higher the variability of Poisson's arrival rates in MMPP, the higher is the clustering of events. Moreover, individual clusters have bigger sizes and smaller time scales, creating short-lived but intense clusters.  

\large To summarize, using standard arrival processes, we illustrated the ability of our complex network framework to characterize an arrival time series.
\section{Real-world applications} \label{Sec: real world applications}

\large In this section we will illustrate our proposed complex network framework by applying it to two real-world applications. Firstly, we will apply the method to a droplet arrival time series in turbulent flows as it represents a compelling example of events characterized by highly irregular time intervals. Here, the time series directly reflects the complex interplay between fluid turbulence and droplet dynamics. Using the proposed method, we will characterize the preferential concentration or spatial clustering of droplets in turbulence. Later, we will analyze the time series of R-R intervals in electrocardiogram (ECG) signals to identify cardiac arrhythmias in patients. In both these disparate systems, our method unravels the change in the underlying mechanism.
\\
%%-------------------------------------
\subsection{Analysis of droplet arrival time series}
%%-------------------------------------
\label{Sec: PDPA time series}

\large Droplet in turbulence rather than being randomly distributed in space, tend to preferentially accumulate in specific regions of the flow organizing into localized spatial clusters. Till now, we used the terms “clustering” and “cluster” to refer to temporal clustering in the irregular time series, we stay consistent with that and use term “spatial clustering” and “preferentially concentration”  exclusively to denote physical clustering of droplets in space. This spatial clustering occurs because of the interaction between the turbulent vortices and the inertia of the droplets. This phenomena is set to significantly enhance the collision rate between droplets and play a critical role during the onset of precipitation in warm clouds. 
\\
\large Several studies have investigated the clustering of droplets in clouds using data obtained from measurement probes mounted on aircraft \cite{chaumat2001droplet, marshak2005small, kostinski2001scale} and on balloons.\cite{lehmann2007evidence} Ideally, capturing the spatial clustering requires higher-dimensional (2D or 3D) in situ measurements such as high-speed laser based imaging of droplets or in-line holography.\cite{fugal2009cloud, larsen2018fine} However, performing these optical measurements is particularly challenging due to the precise alignment of optical systems and the limited space available for equipment in platforms like balloons or aircraft. 
\\
\large Phase Doppler Particle Analyzer (PDPA) or Phase Doppler Anemometer (PDA) is typically used to measure the cloud droplet characteristics such as size and velocity.\cite{albrecht2013laser} In these measurement techniques two laser beams are transmitted which intersect to form a probe volume. When a droplet passes through this probe volume the receiver records the scattered light, which is then processed to obtain the droplet characteristics. Though the instruments monitor the flow continuously, the data acquisition occurs only when a droplet arrives within the probe volume. Hence, the arrival time series of droplets from these measurement techniques are inherently irregular and contain the information of droplet dynamics and the spatial clustering. Thus, analyzing the droplet arrival time series provides insight into the preferential concentration patterns that emerge from droplet–turbulence interactions.
\\
\large Commonly used tools to quantify clustering from time series of droplet arrivals are the fishing statistic \cite{baker2010analysis,uhlig1998holographic}, the clustering index \cite{chaumat2001droplet}, the correlation dimension \cite{shaw2002towards}, and the pair correlation function.\cite{jameson2000fluctuation,larsen2014recovery, saikranthi2013identification} As discussed earlier, these methods typically evaluate the deviations from a Poisson process and provide a global estimate of clustering. However, given the highly multiscale nature of droplet clustering in turbulence, these methods are not suitable for fully characterizing the underlying dynamics. Thus, the developed complex network framework enables exploration of the spatial clustering characteristics of droplets in turbulence through analysis of the droplet arrival time series.

\begin{figure}[ht]
    \centering
    \includegraphics[width=0.75\linewidth]{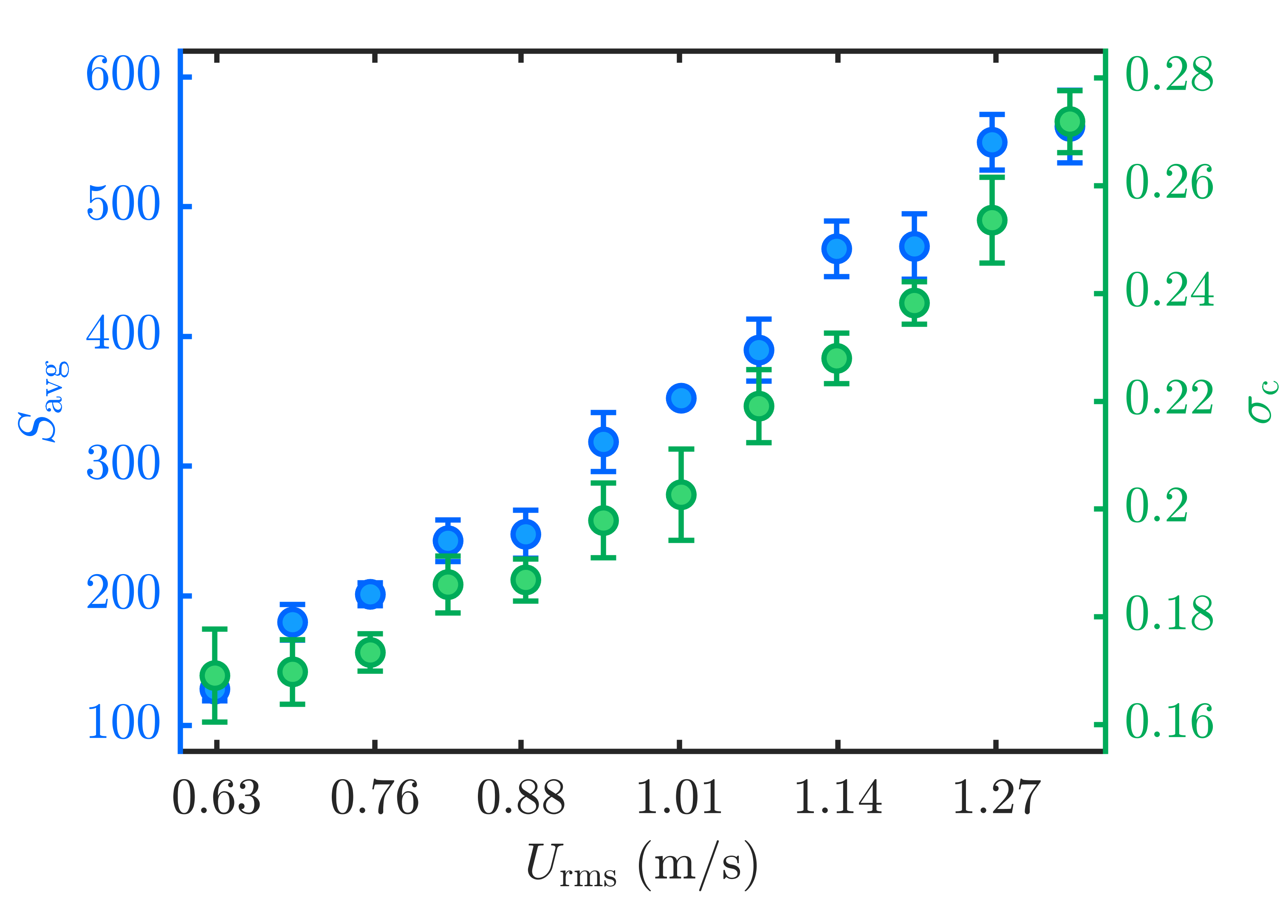}
    \caption{Variation of the average node strength $S_{avg}$ obtained from the droplet arrival time-series and the standard deviation of normalized Voronoi areas $\sigma_{c}$ obtained from spatial measurements as a function of turbulence intensity $U_{rms}$. Both measures increase with turbulence intensity, demonstrating that droplets cluster more at higher turbulence.} 
    \label{Fig: Savgandsigmac}
\end{figure}

\large We analyzed the droplet arrival time series from the laboratory experiments performed in the turbulence chamber facility. Details regarding the experimental setup and measurement techniques are described in detail in Shri Vignesh \textit{et al.}\cite{ShriVigneshTurb} A summary of the experimental conditions and the measurements performed is provided in Appendix~\ref{Apend: B}. We performed a multiscale analysis of droplet arrival time series using the proposed complex network approach.
\\
\large Figure~\ref{Fig: Savgandsigmac} shows the variation of $S_{avg}$ with turbulence intensity $(U_{rms})$. $S_{avg}$ increases with increasing turbulence intensity, indicating enhanced clustering at high turbulence. We compared this result with that obtained using the spatial data from planar Mie-scattering images, and the Voronoi analysis was used to quantify the spatial clustering.\cite{obligado2014preferential, monchaux2010preferential, ferenc2007size} Voronoi analysis quantifies droplet clustering from images by dividing the image into Voronoi cells around each droplet, allowing the local cell area distribution to serve as a statistical measure of clustering intensity.\cite{monchaux2010preferential} The degree of clustering is quantified using the clustering contribution of the standard deviation of the normalized Voronoi cell areas $\sigma_{c}$ following the method described by Sumbekova \textit{et al.}\cite{sumbekova2017preferential} Both measures increase with turbulence intensity, indicating that the average node strength obtained from the time series follows a trend similar to that of the spatial clustering measure. This comparison demonstrates that the proposed complex network approach can reliably estimate spatial clustering from one-dimensional droplet arrival measurements.

\begin{figure}
    \centering
    \includegraphics[width=0.75\linewidth]{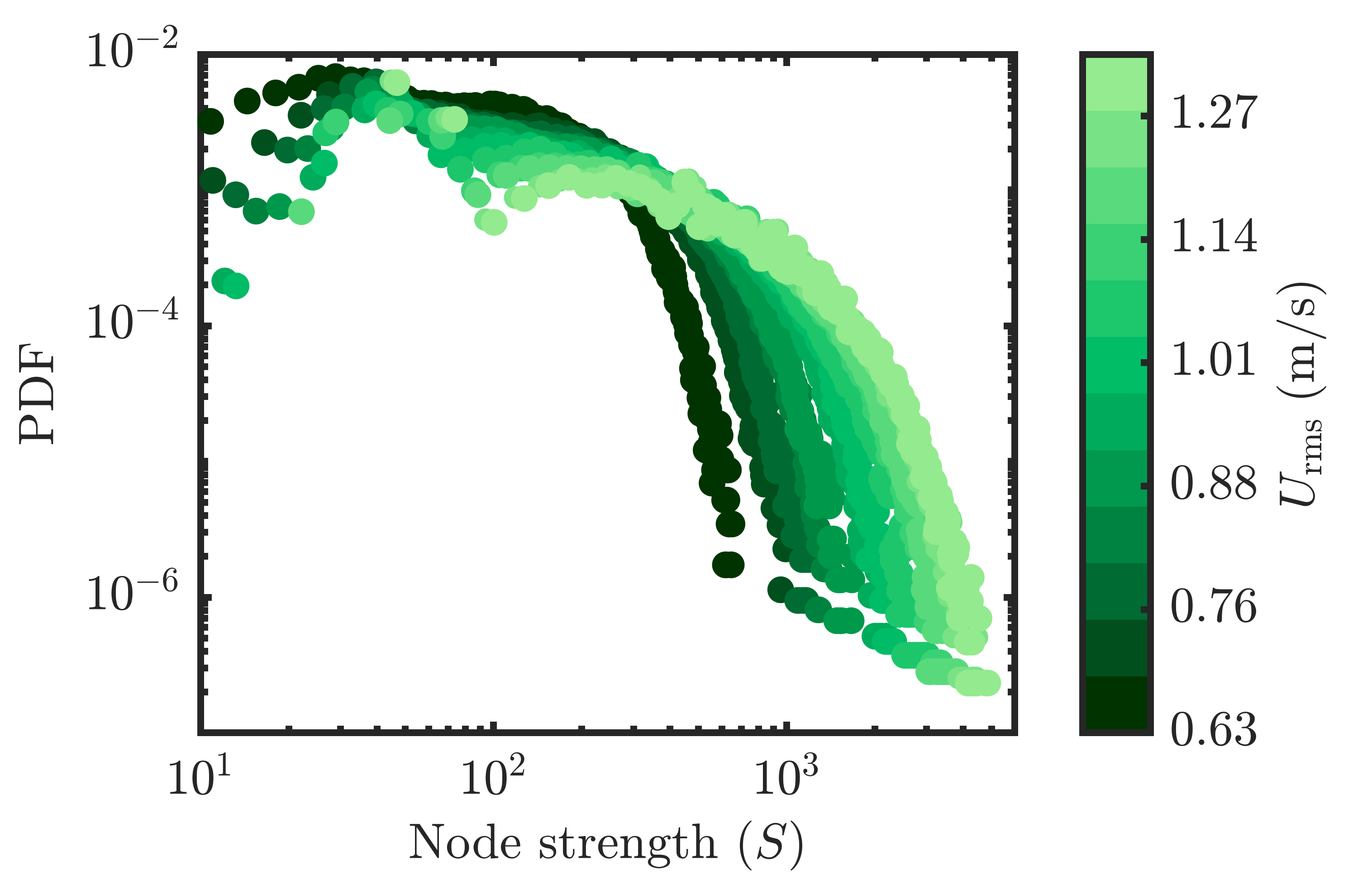}
    \caption{PDF of node strength ($S$) of the droplets based on their arrivals. We observe a significant deviation from the distribution observed for Poisson arrivals, signifying enhanced clustering. Furthermore, the presence of highly clustered nodes with increasing $U_{rms}$ indicates highly clustered droplets occurring at close intervals.}
    \label{Fig: PDF_of_NS}
\end{figure}

\begin{figure}
    \centering
    \includegraphics[width=0.85\linewidth]{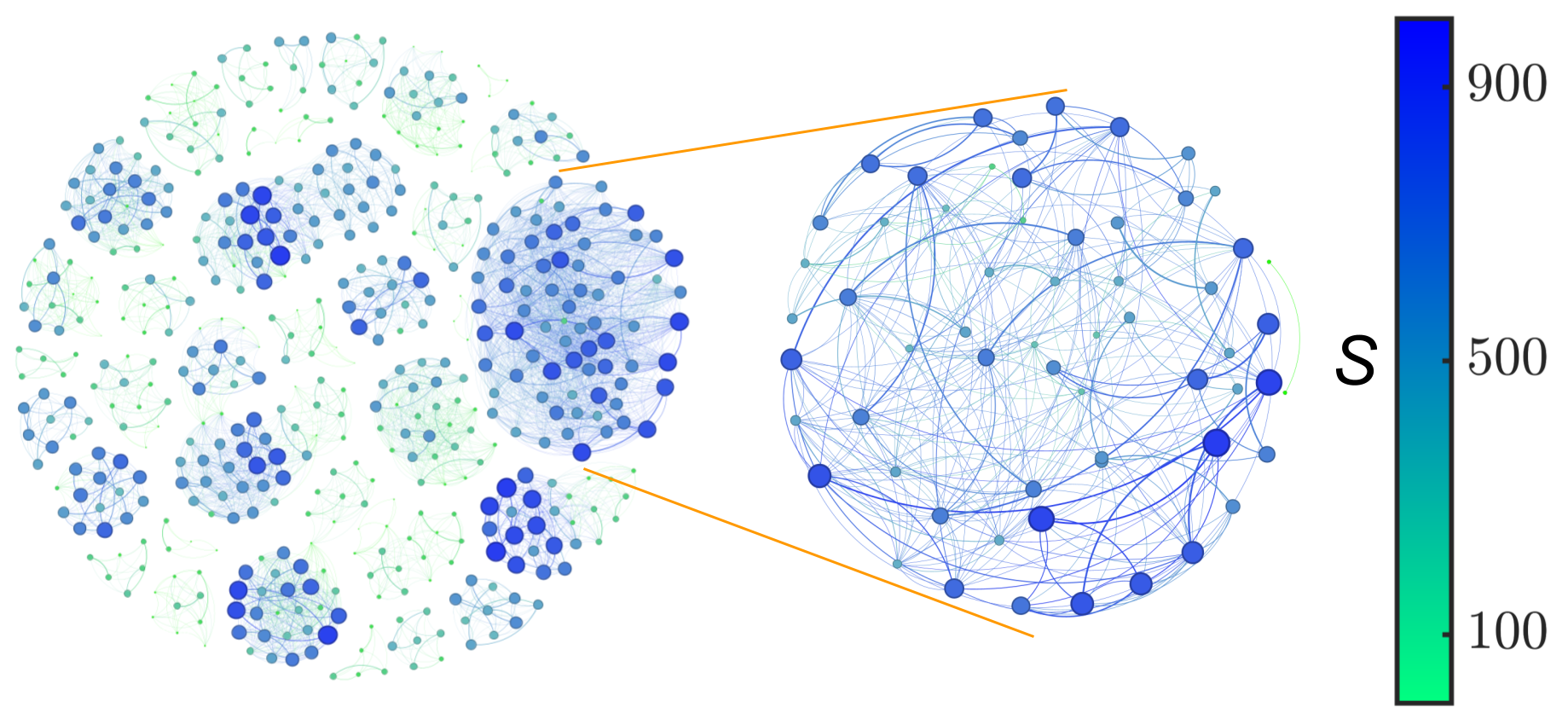}
    \caption{Network topology for droplet arrival time series at $U_{rms}=1.01 \text{\ m/s}$ plotted using Fruchterman and Reingold algorithm\cite{fruchterman1991graph} in Gephi visualization software \cite{bastian2009gephi}. The topology of the network shows distinct clusters with many intra-cluster connections but fewer inter-cluster links. The figure on the right shows the network for one of the clusters.}
    \label{Fig: Network}
\end{figure}

\large The node strength ($S$) distribution reflects the local clustering experienced by a droplet during its arrival (Fig.~\ref{Fig: PDF_of_NS}). A clear distinction emerges between the distribution of Poisson arrivals and that of droplets in turbulence. Droplets exhibit a broad range of node strengths, spanning from $10$ to $10^3$, in contrast to the narrower distribution earlier observed for Poisson arrivals (Fig.~\ref{Fig: Node strength} (b)). Moreover, arrivals with very high node strengths are absent in the Poisson case. These highly clustered droplet arrivals are particularly significant, as they may enhance collision and coalescence processes, thereby influencing the overall dynamics of droplet–turbulence interactions.
\\
\large While the quantitative comparison with spatial clustering demonstrates the ability of the network-based approach to estimate clustering, visualization of the constructed networks provides additional insight into the underlying structure of clustering. Network visualization allows inspection of how arrivals group together and how individual clusters are separated. Figure~\ref{Fig: Network} shows the network graph corresponding to a segment of droplet arrival time series at $U_{rms} = 1.01 \text{\ m/s}$. We observe the formation of distinct groups of nodes, indicating the presence of well-defined clusters. The zoomed-in view of a cluster (right) reveals high number of links within the cluster, in contrast to the relatively sparse connections between nodes belonging to different clusters.

\large We apply a community detection algorithm to the network constructed from the droplet arrival time series. The communities identified in the network correspond to clusters of temporally grouped droplet arrivals in the time series. To avoid spurious detections, communities containing fewer than three droplet arrivals are excluded from further analysis. The remaining communities are treated as individual droplet clusters for further analysis.

\begin{figure}
    \centering
    \includegraphics[width=0.75\linewidth]{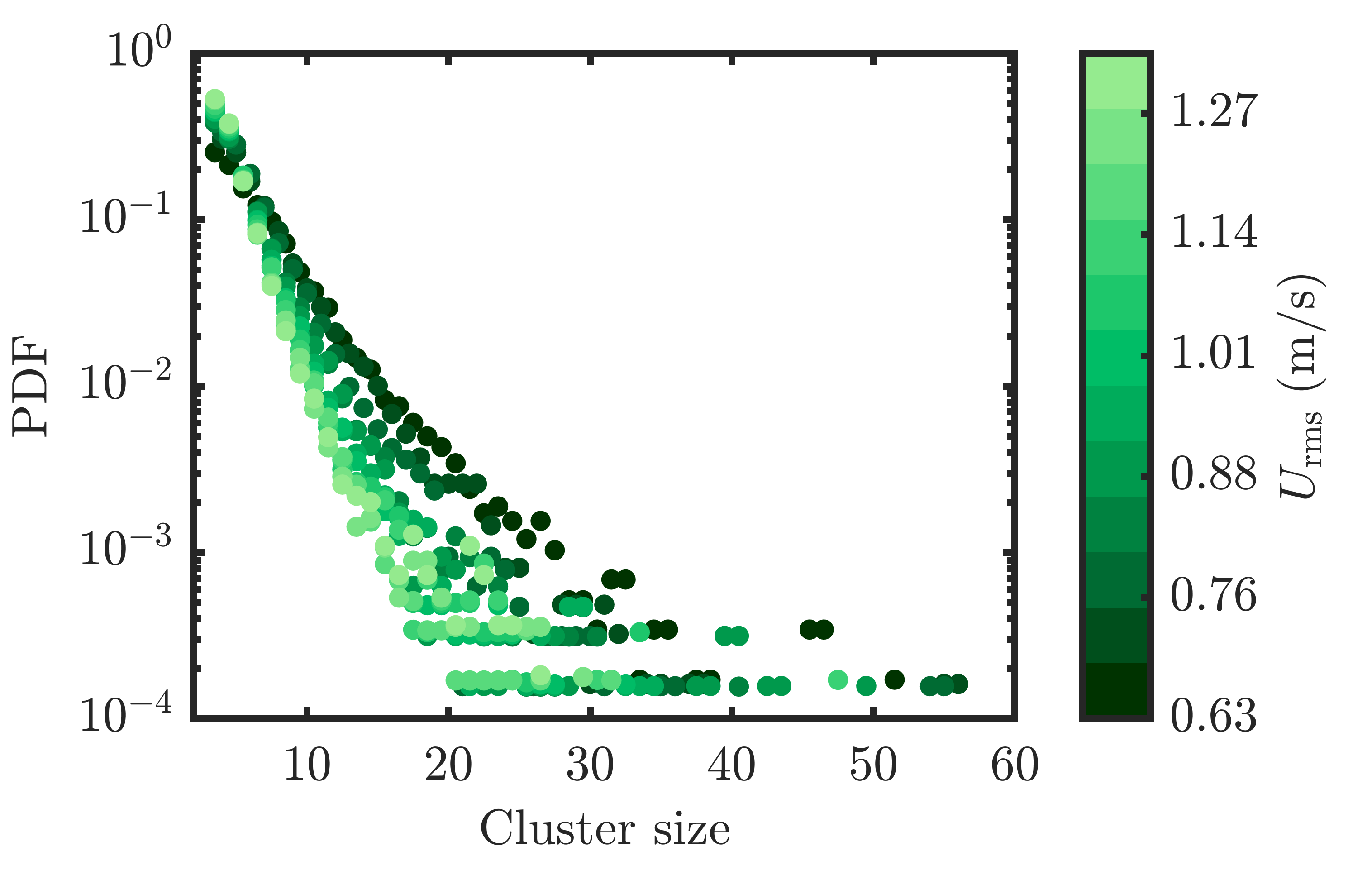}
    \caption{PDF of the number of arrivals (size of droplet clusters). We observe that the cluster size decreases with increasing $U_{rms}$. This trend suggests that enhanced turbulence intensities promote the formation of dense clusters with fewer droplets.}
    \label{Fig: PDF_clu_size}
\end{figure}

\large Figure~\ref{Fig: PDF_clu_size} shows PDF of the number of arrivals in individual cluster for different $U_{rms}$. With increasing $U_{rms}$ the number of droplets in the arrivals decreases, possibly leading to the formation of denser clusters containing fewer droplets. This trend suggests that stronger turbulence enhances droplet mixing and dispersal, breaking up larger clusters into smaller, more compact ones. Consequently, the clustering becomes more localized, reflecting the intensified influence of turbulent eddies.
\\
\large For each identified cluster, we examine the distribution of droplet diameters and quantify the variability in droplet size using the coefficient of variation $(C_{v})$. The coefficient of variation is defined as
\begin{equation}\label{Eq: coefficient of variation }
    \begin{aligned}
        C_{v}=\frac{\sigma_{d}}{\mu_{d}},
    \end{aligned}
\end{equation}
where $\mu_{d}$ and $\sigma_{d}$ denote the mean and standard deviation of droplet diameters within a cluster, respectively. The $C_{v}$ provides a normalized measure of variability \cite{abdi2010coefficient}, with lower values indicating relatively uniform droplet sizes within a cluster and higher values indicating greater size variation.

\begin{figure}
    \centering
    \includegraphics[width=0.75\linewidth]{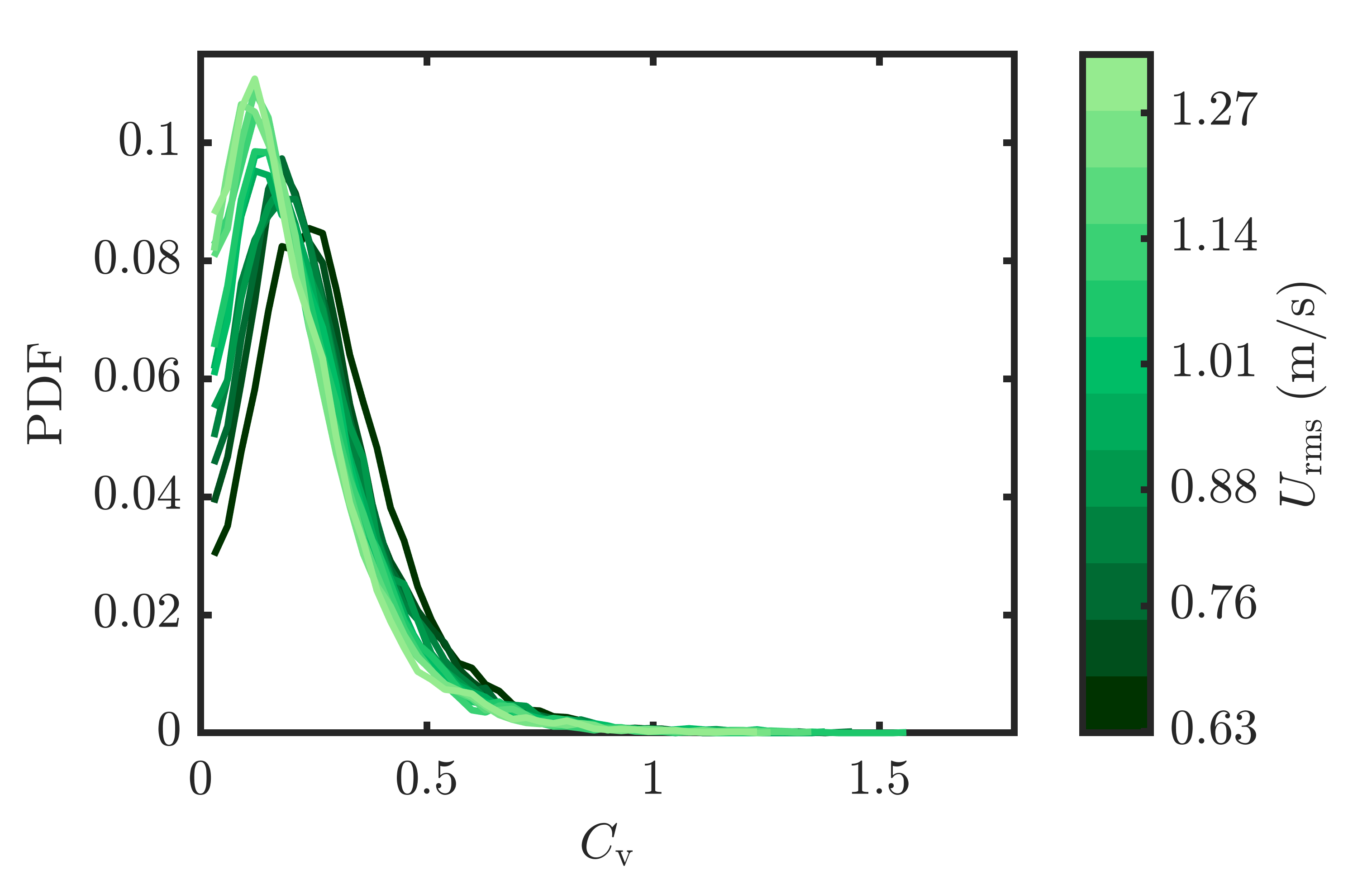}
    \caption{PDF of the coefficient of variation ($C_{\mathrm{v}}$) of droplet diameters computed for all identified temporal clusters across all turbulence intensities. The distribution shifts towards smaller values of $C_{\mathrm{v}}$ with increase in $U_{\mathrm{rms}}$ indicating a greater coherence in droplet size within clustered droplets.}
    \label{Fig: Cv_F}
\end{figure}

\large Figure~\ref{Fig: Cv_F} shows the distribution of $C_{v}$ values computed across all identified clusters for each turbulence intensity. The distribution is strongly skewed toward low $C_{v}$ values, indicating that, for most clusters, the droplet within a cluster are relatively similar in size. This observation suggests that droplets grouped together in the arrival time series are not only clustered temporally but also exhibit coherence in droplet size. Such size homogeneity within identified clusters points to an underlying organization in the droplet arrival process that extends beyond purely temporal correlations. Furthermore, the distribution shifts toward smaller values of $C_{v}$ with increasing $U_{rms}$, indicating higher coherence in the size of clustered droplets. This trend suggests that stronger turbulence promotes more uniform droplet sizes inside clusters.

\begin{figure}
    \centering
    \includegraphics[width=0.75\linewidth]{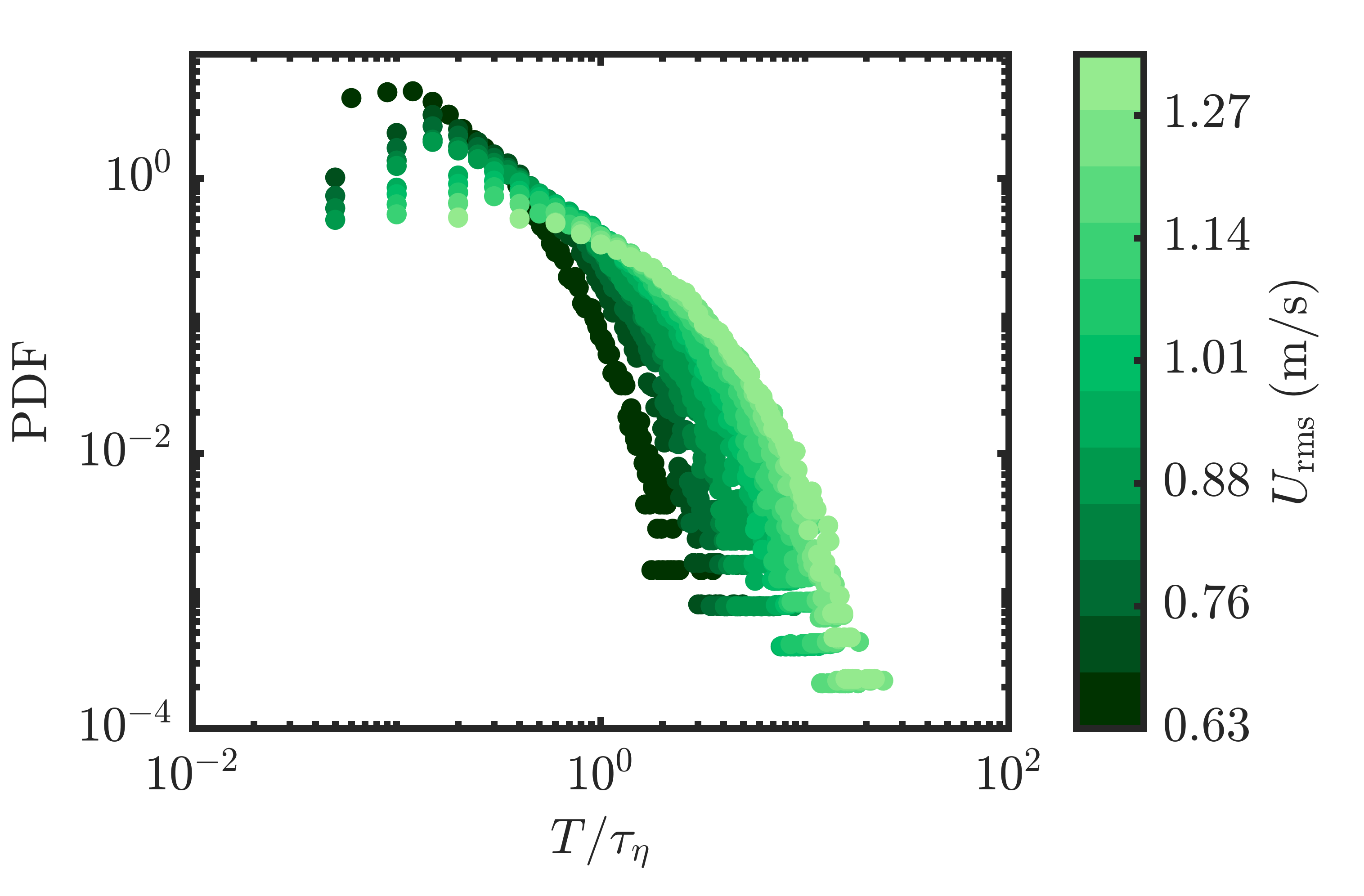}
    \caption{PDF of the normalized cluster time scales ($T/\tau_{\eta}$) reveals the presence of clusters spanning a wide range of temporal variability. As $U_{\mathrm{rms}}$ increases, the droplet clusters become more persistent relative to $\tau_{\eta}$.} 
    \label{Fig: PDF_of_Time_scale}
\end{figure}

\large The cluster time scale ($T$), or time span, is defined as the duration between the first and last arrivals within a given cluster. We normalize the time scale using the Kolmogorov time scale $\tau_{\eta}$. Figure ~\ref{Fig: PDF_of_Time_scale} shows the PDF of normalized cluster time scale ($T/\tau_{\eta}$) for different $U_{rms}$. We observe droplet clusters to occur in a wide range of temporal variability. Furthermore, with increasing $U_{rms}$ we observe a shift in the distribution towards longer-lived droplet clusters (i.e., larger time scale).

\subsection{Detecting heart arrhythmia}
\label{Sec: Heart}

\begin{figure}
    \centering
    \includegraphics[width=0.75\linewidth]{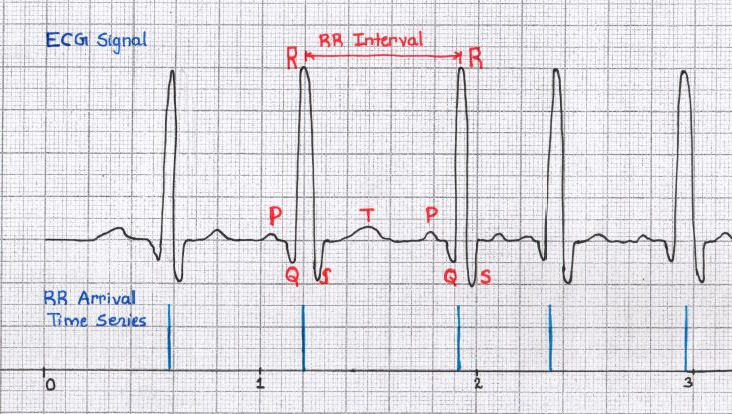}
    \caption{ECG signal with the P wave, QRS complex, T wave, and RR interval. Considering R peak as arrival gives RR arrival time series. The time between two R peaks is not constant, and hence, we get an irregular arrival time series, as shown in the bottom panel. The vertical lines depict RR arrivals.}
    \label{Fig: ECG}
\end{figure}

\large In the analysis presented above, we consider systems in which the time series data have already been recorded and are available for offline analysis. However, in many real-world systems, the underlying dynamics evolve continuously in time, and therefore, the data must be processed in real time as new observations arrive. A commonly adopted strategy for such real-time analysis is the moving-window (or sliding-window) approach \cite{Kantz2004}, which has been widely employed in natural \cite{dakos2024tipping,yang2022critical,pace2017reversal} and engineering \cite{banerjee2024early,radhakrishnan2025early} systems to extract evolving features and to identify early warning signals. In this approach, a measure is computed from the most recent segment of the time series within a finite window, and the window is then progressively shifted forward as new data points become available. This procedure enables continuous monitoring of the system dynamics over time. In the present work, we apply the proposed complex-network framework within a moving-window scheme to analyze RR-interval time series obtained from electrocardiogram (ECG) recordings. This allows the network-based clustering measure to be computed dynamically from successive segments of the data, thereby enabling real-time assessment of changes in heartbeat dynamics.
\\
\large The behavior of heartbeats has been extensively studied over several decades. An electrocardiogram (ECG) is a widely used tool for recording the electrical activity of the heart and for diagnosing and monitoring various cardiac conditions. The ECG signal contains distinct waves and intervals that correspond to the electrical activity of the atria and ventricles. A representative ECG signal showing these waves and intervals is presented in Fig.~\ref{Fig: ECG}. An ECG signal consists of the P wave, QRS complex, T wave, PR interval, QT interval, and ST segment.

\begin{figure}
    \centering
    \includegraphics[width=0.75\linewidth]{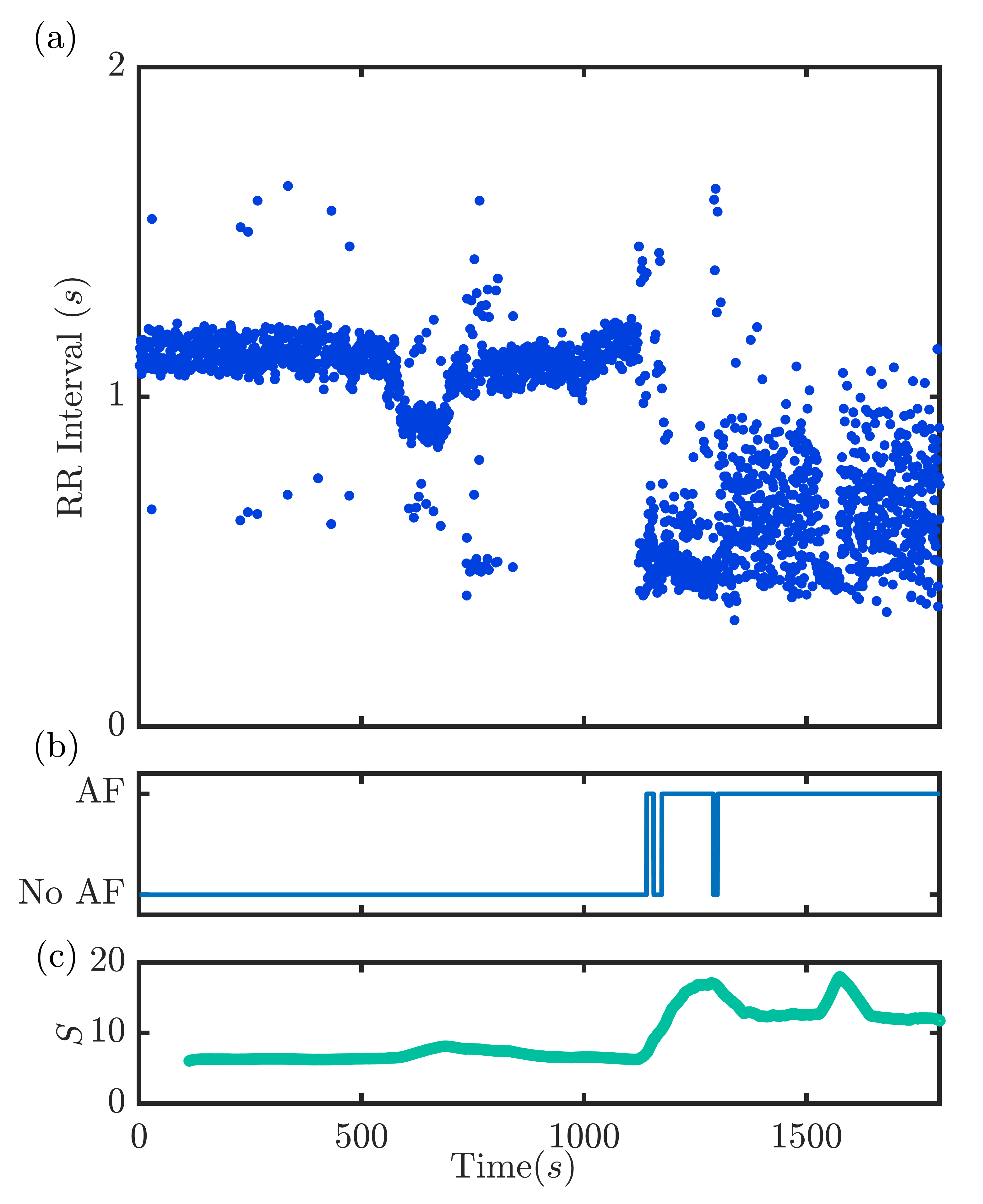}
    \caption{\textbf{(a)} The time series of RR intervals for record 202 from MIT-BIH arrhythmia database, \textbf{(b)} atrial fibrillation episodes annotated by experts, \textbf{(c)} the node strength variation calculated for the RR arrival time series. The atrial fibrillation episodes marked by experts are towards the end of the signal, where the RR interval value reduces. The node strength value captures the clustering of events in RR arrivals and hence increases during episodes of atrial fibrillation.} 
    \label{Fig: HR_AF detection}
\end{figure}
\large The RR interval, defined as the time between successive R peaks, exhibits continuous variability, resulting in an irregular sequence of R-peak arrivals. Time series of RR intervals have been widely used for the analysis and classification of cardiac arrhythmias.\cite{carrara2015heart} Atrial fibrillation is one of the most common arrhythmias and is associated with irregular atrial activation. During atrial fibrillation, the atrial activations are irregularly relayed through the atrioventricular node, resulting in a correspondingly irregular sequence of ventricular activations, which can be captured through RR intervals.\cite{tateno2001automatic}
\\
\large RR intervals are inherently nonuniform and form an irregular arrival-time series. In this work, we analyze the arrival times of RR intervals extracted from ECG signals to detect atrial fibrillation. The RR arrival data are obtained from the MIT–BIH Arrhythmia Database available on PhysioNet.\cite{goldberger2000physiobank} This database contains 48 half-hour excerpts of ECG recordings obtained from 47 subjects. In addition to the ECG signals and corresponding RR-interval time series, the database provides beat annotations. These annotations are labels assigned by experts that indicate specific locations within the signal and describe the cardiac events occurring at those points. The beat annotations classify the signals into normal sinus rhythm, atrial fibrillation, and other arrhythmias.\cite{moody2001impact}
\\
\large The time series of RR intervals for record 202, along with expert-annotated atrial fibrillation episodes, are shown in Fig.~\ref{Fig: HR_AF detection} \textbf{(a)-(b)}. During atrial fibrillation episodes, the RR intervals become shorter on average and exhibit increased variability, reflecting the irregular transmission of atrial activity to the ventricles. The increased variability in RR intervals can give rise to clustering in the corresponding RR arrival time series. To quantify this behavior, we analyze the RR arrival-time series using the proposed network-based framework and compute the node strength as a measure of event clustering. The temporal evolution of node strength is shown in Fig.~\ref{Fig: HR_AF detection} \textbf{(c)}. The node strength is evaluated within a moving window of 30 seconds along the RR arrival sequence, allowing the clustering measure to be updated continuously as new events arrive. A clear increase in node strength is observed during atrial fibrillation episodes identified by expert annotations (Fig.~\ref{Fig: HR_AF detection} \textbf{(b)}), indicating enhanced clustering of RR arrivals during these intervals.
\\
\large These results demonstrate that the network-based clustering measure is sensitive to changes in the temporal organization of heartbeats and can effectively distinguish atrial fibrillation episodes from normal sinus rhythm. Unlike conventional heart rate variability measures, which typically require the complete ECG signal and rely primarily on global statistics, the proposed approach captures clustering directly at the level of event arrivals. Hence, the method enables the identification of rhythm irregularities using only RR arrival information and can be naturally implemented for real-time monitoring of cardiac dynamics. These findings highlight the potential of the proposed framework as a tool for detecting atrial fibrillation and for providing timely indications of arrhythmia-related changes in cardiac rhythm.
%%=========================================
\section{Discussion and Conclusion} \label{Sec: Conclusions}
%%=========================================
\large The temporal inhomogeneity inherent in irregular time series gives rise to clustering of events, reflecting the underlying dynamical mechanisms governing the system evolution. In this work, we presented a complex network–based framework to analyze clustering in irregular time series. The proposed approach enables multiscale characterization by quantifying clustering at both the event and global levels while simultaneously identifying and characterizing individual clusters.
We first validated the framework using standard arrival processes, such as regular arrivals, Poisson processes, and Markov-modulated Poisson processes (MMPP). As expected, regular arrivals exhibit no clustering, Poisson processes display low clustering, and MMPP time series show pronounced clustering. The network representation clearly reflects these differences: MMPP networks display well-defined communities, whereas regular and Poisson arrivals lack such organized structures. Systematic variation of MMPP parameters further demonstrates that increasing the variability of the underlying Poisson rate produces larger and stronger, yet temporally shorter-lived clusters, highlighting the role of rate modulation in shaping clustering behavior.
\\ 
\large We then applied the clustering analysis to time series of droplet arrivals acquired from experiments in a turbulence chamber. Clustering quantified using the network-based measure $S_{avg}$ exhibits trends consistent with the spatial clustering measure $\sigma_{c}$ obtained from Voronoi tessellation of Mie-scattering images. This agreement demonstrates that spatial preferential concentration can be reliably inferred from one-dimensional droplet arrival measurements. The analysis further shows that droplet clustering increases with turbulence intensity. Beyond reproducing global trends, the network framework reveals additional structure. Community detection identifies distinct droplet clusters, enabling analysis of their internal properties and temporal scales. Droplets within a given temporal cluster exhibit low size variability, indicating coherence in droplet properties that extends beyond mere temporal proximity. Additionally, the number of droplet arrivals within individual clusters decreases with increasing turbulence intensity, suggesting the formation of more compact and strongly clustered structures. Cluster lifetimes span a broad range of scales and shift toward longer durations at higher turbulence intensities, highlighting the inherently multiscale nature of droplet–turbulence interactions.
\\ 
\large Finally, we analyzed clustering in the time series of RR intervals using the proposed network method. The results show that node strength increases during atrial fibrillation episodes, reflecting enhanced clustering in RR arrivals. This enables the identification of atrial fibrillation from RR arrival time series alone.
\\ 
\large Overall, while droplet arrival time series and ECG data are used here as representative examples, the proposed framework is broadly applicable to irregular time series arising from diverse systems. By combining node-level metrics, community detection, and cluster-scale analysis, the approach simultaneously captures local clustering intensity, cluster coherence, and system-wide organization.

\section{Acknowledgments}
We thank Ms. Sruthibhai, Ms. Arya, Ms. Sudha, Mr. Thilagaraj S., and Mr. Anand S. for their technical support in experiments. S. S. Ambedkar and K. S. Vignesh are thankful to the Ministry of Education (MoE) for the HTRA. The work on detecting heart arrhythmia (Section \ref{Sec: Heart}) was carried out with support from IOE (Project no. SP22231222CPETWOCTSHOC). R. I. Sujith thanks the ISRO-IIT(M) cell (Project No. SP/21-22/1197/AE/ISRO/002696) for funding the research on droplet arrival time series data presented in Section \ref{Sec: PDPA time series}

\section{Author Declarations}
\textbf{Conflict of Interest:}
The authors have no conflicts to disclose.
\\
\textbf{Data availability:}
The data that support the findings of this study are available
from the corresponding author upon reasonable request.

\section{Appendix} \label{Sec: Appendix}

\subsection{Generation of arrival time series for each arrival process.}
\label{Apend: A}

\large We generate a time series with 10000 arrivals at the average arrival rate of 1000 arrivals per second  ($AR = 1000$). For regular arrival, the interarrival time remains constant and is given as $t = 1/AR$. Let $T_{i}$ be the array of arrival times. The arrivals are simulated by taking the first arrival at time $t$, i.e. $T_{1}=t$. The consecutive arrivals are then computed by adding $t$ to the previous arrival, 
\begin{equation}\label{Eq: Reg arrival}
    \begin{aligned}
        T_{i+1} =T_{i} + t.  
    \end{aligned}
\end{equation}
For Poisson's arrival process, the arrival rate is given as $\lambda$; for our case, $\lambda = AR$. The interarrival times are exponentially distributed in Poisson's arrival process.\cite{daw2018queues} We generate an array $(t_{i})$ of exponentially distributed random numbers with mean $\mu = 1/\lambda$. This array corresponds to interarrival times; let $t_{i}=[t_{1},t_{2} ..... t_{n}]$ be the array of interarrival time. The first arrival is scheduled at $t_{1}$, i.e., $T_{1}=t_{1}$, and consecutive arrival times are calculated by adding a value of time interval from the interarrival time array to the previous arrival. The arrival time for Poisson's arrival process is given as,
\begin{equation}\label{Eq:poi arrival}
    \begin{aligned}
        T_{i+1} =T_{i} + t_{i+1}.  
    \end{aligned}
\end{equation}

\large Markov-modulated Poisson process can be constructed by varying the arrival rate of a Poisson's arrival process according to an \textit{m}-state Markov chain, which is independent of the arrivals.\cite{fischer1993markov} When the Markov chain is in state \textit{i}, arrivals occur according to Poisson's process with arrival rate $\lambda_{i}$. The MMPP is parameterized by the \textit{m}-state continuous-time Markov chain with infinitesimal generator\cite{cinlar1975introduction} $Q$ and m Poisson's arrival rates $\lambda_{1}$,$\lambda_{2}$,....,$\lambda_{m}$. where,

\begin{equation}\label{Eq: infinitesimal generator 1}
    \begin{aligned}
        Q=  \begin{bmatrix}
            -\sigma_{1}  &   \sigma_{12}  & \cdots    & \sigma_{1m}\\
            \sigma_{21}  &   -\sigma_{2}  & \cdots    & \sigma_{2m}\\
            \vdots  &  \vdots   & \ddots    & \vdots\\
            \sigma_{m1}  &   \sigma_{\mathrm{m2}}  & \cdots    & -\sigma_{m}\\
             \end{bmatrix},
     \end{aligned}
\end{equation}

\begin{equation}\label{Eq: infinitesimal generator 2}
    \begin{aligned}
        \sigma_{i} = \sum_{{j=1}}^{m} \sigma_{ij} , \qquad  j \neq i  ,  
    \end{aligned}
\end{equation}

\begin{equation}\label{Eq: Lambda 1}
    \begin{aligned}
        \Lambda= \begin{bmatrix} 
                    \lambda_{1},  &   \lambda_{2},  & \cdots    & \lambda_{m}
                 \end{bmatrix}  .
    \end{aligned}
\end{equation}
We generate MMPP arrival time series from \textit{m}-state continuous time Markov chain where m=100 and infinitesimal generator $Q$ computed from $\sigma_{i}$ given as,
\begin{equation}\label{Eq: Lambda 2}
    \begin{aligned}
        \sigma_{\mathrm{ij}} = \left\{ \begin{array}{ll}
                      -1, & \qquad i=j\\
                      \frac{1}{m-1}, & \qquad  i \neq j 
                      \end{array}\right.
    \end{aligned}
\end{equation}

\large We generate 100 random values for the Poisson's arrival rate array ($\Lambda$) with a mean of 1000 and standard deviation of 100, i.e. $\Lambda_{m}=1000$ and $\Lambda_{sd} = 100$. The transition probability matrix \cite{fischer1993markov,ryden1996algorithm} is then computed using $Q$ and $\Lambda$. The first 100 arrivals are generated as Poisson's arrival with an arrival rate of $\lambda_{1}$; the next arrival rate is obtained using transition probabilities, and 100 more arrivals are generated using this arrival rate. This procedure is continued 100 times, yielding 10000 Markov-modulated Poisson process arrivals. 
\\
\large Further, we create five different MMPP arrival time series to implement community detection and study individual clusters. Different MMPP arrival time series are created by changing the $\Lambda$ array and keeping all other parameters the same. We keep $\Lambda_{m}$ constant and equal to 1000 and change $\Lambda_{sd}$ from 100 to 300 in steps of 50 where case 1 has $\Lambda_{sd}$ = 100 and case 5 has $\Lambda_{sd}$ = 300. With the increase in $\Lambda_{sd}$, we increase the variability in Poisson's arrival rate ($\lambda$), giving rise to arrivals with different arrival rates. By doing this, we attempt to vary the clustering of MMPP arrivals and test the ability of our method to identify individual clusters.

\subsection{Experimental setup for acquiring droplet arrival data}
\label{Apend: B}

\begin{figure}[h]
    \centering
    \includegraphics[width=0.85\linewidth]{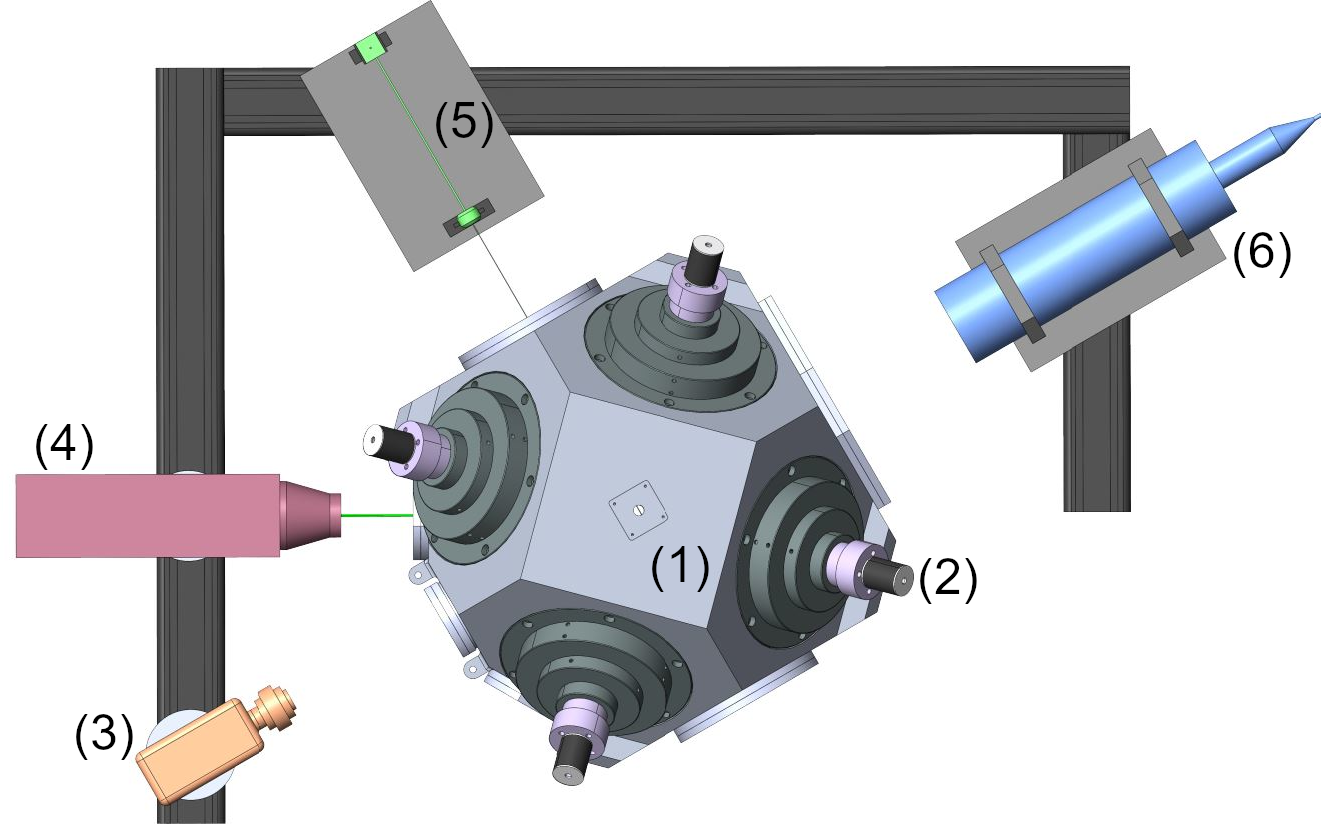}
    \caption{Schematic of experiment setup in top view with (1) turbulence chamber, (2) servo motor, (3) high-speed camera (4) PDPA transmitter (5) laser light sheet and (6) PDPA receiver probe } 
    \label{Fig: Exp.Setup}
\end{figure}

\begin{figure}
    \centering
    \includegraphics[width=0.75\linewidth]{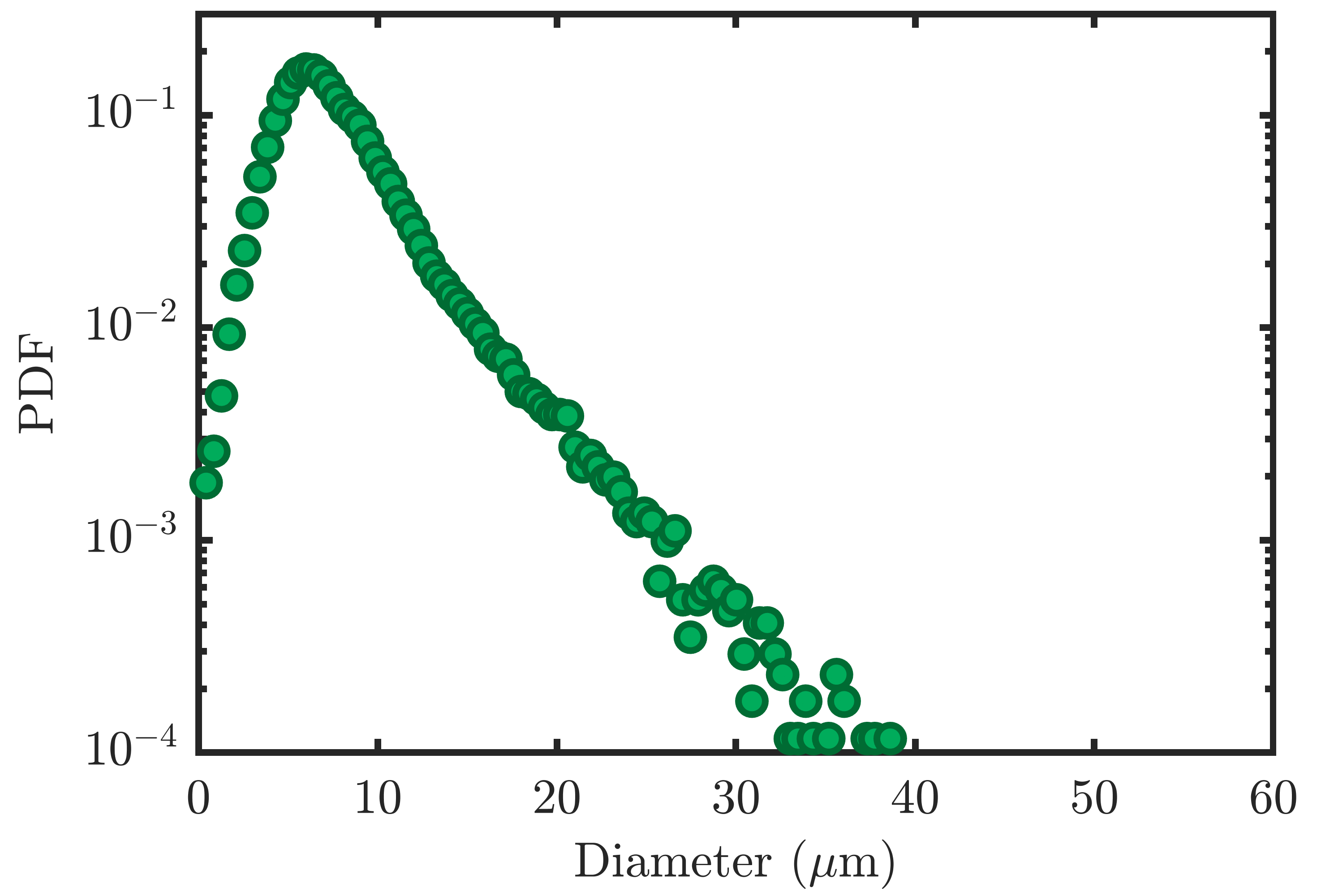}
    \caption{PDF of droplet size at lowest turbulence intensity ($U_{rms} = 0.63 \text{\ m/s}$)} 
    \label{Fig: Dsd}
\end{figure}

Droplet-turbulent interactions are studied in our turbulence chamber facility shown in the schematic Fig.~\ref{Fig: Exp.Setup}. The turbulence chamber is a spherical cavity equipped with eight symmetrically mounted servo motors, each fitted with a fan, designed to generate homogeneous and isotropic turbulence. Turbulence intensity is controlled by controlling the fan speed. The droplets are seeded into the chamber using microporous atomizers. Figure~\ref{Fig: Dsd} shows the PDF of the droplet size at the lowest turbulence intensity.

For each experimental condition, droplet size and velocity were measured using a Phase Doppler Particle Analyzer (PDPA), and planar Mie-scattering images of the droplets were subsequently obtained. A Nd:YLF laser operating at 527 $nm$, with a maximum energy of 25 $mJ$ per pulse at 1 $kHz$, was used to generate the laser beam. The droplets passing through this laser sheet scatter the laser light, which are captured using a high-speed imaging camera with a spatial resolution $1280 \times 1024$ at 23 $\mu$m per pixel. The Mie-scattering image reveals the position of droplets in the plane of the laser sheet, giving the 2D spatial distribution of droplets. Experiments were performed for 12 different turbulence intensities, and for each experimental condition, 4 independent realizations were performed. In each experiment, 10,000 droplet arrivals were recorded using PDPA, and 2,000 images were acquired.

\bibliographystyle{unsrtnat}

\bibliography{references}

@PREAMBLE{
 "\providecommand{\noopsort}[1]{}" 
 # "\providecommand{\singleletter}[1]{#1}%" 
}

@article{daw2018queues,
  title={Queues driven by {H}awkes processes},
  author={Daw, Andrew and Pender, Jamol},
  journal={Stochastic Systems},
  volume={8},
  number={3},
  pages={192--229},
  year={2018},
  publisher={INFORMS},
  doi={10.1287/stsy.2018.0014},
  URL={https://doi.org/10.1287/stsy.2018.0014}
}

@article{ogata1988statistical,
  title={Statistical models for earthquake occurrences and residual analysis for point processes},
  author={Ogata, Yosihiko},
  journal={Journal of the American Statistical Association},
  volume={83},
  number={401},
  pages={9--27},
  year={1988},
  publisher={Taylor \& Francis},
  doi={10.1080/01621459.1988.10478560},
  URL={https://www.tandfonline.com/doi/abs/10.1080/01621459.1988.10478560}
}

@article{reno2003closer,
  title={A closer look at the {E}pps effect},
  author={Ren{\`o}, Roberto},
  journal={International Journal of Theoretical and Applied Finance},
  volume={6},
  number={01},
  pages={87--102},
  year={2003},
  publisher={World Scientific},
  doi={10.1142/S0219024903001839},
  URL={https://doi.org/10.1142/S0219024903001839}
}

@article{azizpour2018exploring,
  title={Exploring the sources of default clustering},
  author={Azizpour, Shahriar and Giesecke, Kay and Schwenkler, Gustavo},
  journal={Journal of Financial Economics},
  volume={129},
  number={1},
  pages={154--183},
  year={2018},
  publisher={Elsevier},
  doi={https://doi.org/10.1016/j.jfineco.2018.04.008},
  URL={https://www.sciencedirect.com/science/article/pii/S0304405X1830103X}
}

@inproceedings{zhang2004model,
  title={A model-based clustering for time-series with irregular interval},
  author={Zhang, Xiao-Tao and Zhang, Wei and Xiong, Xiong and Wang, Qi-Wen and Li, Cui-Yu},
  booktitle={Proceedings of 2004 International Conference on Machine Learning and Cybernetics},
  pages={2883--2888},
  year={2004},
  organization={IEEE},
  doi = {10.1109/ICMLC.2004.1378524},
  URL = {https://ieeexplore.ieee.org/abstract/document/1378524}
}

@article{telesca2007time,
  title={Time-clustering of natural hazards},
  author={Telesca, Luciano},
  journal={Natural Hazards},
  volume={40},
  pages={593--601},
  year={2007},
  publisher={Springer},
  doi={10.1007/s11069-006-9023-z},
  URL={https://doi.org/10.1007/s11069-006-9023-z}
}

@article{baker1992turbulent,
  title={Turbulent entrainment and mixing in clouds: {A} new observational approach},
  author={Baker, Bradley A},
  journal={Journal of Atmospheric Sciences},
  volume={49},
  number={5},
  pages={387--404},
  year={1992},
  doi = {10.1175/1520-0469(1992)049<0387:TEAMIC>2.0.CO;2},
  url = {https://journals.ametsoc.org/view/journals/atsc/49/5/1520-0469_1992_049_0387_teamic_2_0_co_2.xml}
}

@article{shaw2002towards,
  title={Towards quantifying droplet clustering in clouds},
  author={Shaw, RA and Kostinski, AB and Larsen, Michael L},
  journal={Quarterly Journal of the Royal Meteorological Society: A Journal of the Atmospheric Sciences, Applied Meteorology and Physical Oceanography},
  volume={128},
  number={582},
  pages={1043--1057},
  year={2002},
  publisher={Wiley Online Library},
 doi = {https://doi.org/10.1256/003590002320373193},
 url = {https://rmets.onlinelibrary.wiley.com/doi/abs/10.1256/003590002320373193},
}

@article{baker2010analysis,
  title={Analysis of tools used to quantify droplet clustering in clouds},
  author={Baker, Brad and Lawson, R Paul},
  journal={Journal of the Atmospheric Sciences},
  volume={67},
  number={10},
  pages={3355--3367},
  year={2010},
  doi = {10.1175/2010JAS3409.1},
  url = {https://journals.ametsoc.org/view/journals/atsc/67/10/2010jas3409.1.xml}
}

@incollection{geva2017brain,
  title={Brain state identification and forecasting of acute pathology using unsupervised fuzzy clustering of EEG temporal patterns},
  author={Geva, Amir B and Kerem, Dan H},
  booktitle={Fuzzy and neuro-fuzzy systems in medicine},
  pages={19--68},
  year={2017},
  publisher={CRC Press},
  doi={10.1201/9780203713419-4},
  url={https://www.taylorfrancis.com/chapters/edit/10.1201/9780203713419-4/brain-state-identification-forecasting-acute-pathology-using-unsupervised-fuzzy-clustering-eeg-temporal-patterns-amir-geva-dan-kerem}
}

@article{marwan2015complex,
  title={Complex network based techniques to identify extreme events and (sudden) transitions in spatio-temporal systems},
  author={Marwan, Norbert and Kurths, J{\"u}rgen},
  journal={Chaos: An Interdisciplinary Journal of Nonlinear Science},
  volume={25},
  number={9},
  pages={097609},
  year={2015},
  publisher={AIP Publishing},
  doi={10.1063/1.4916924},
  url={https://doi.org/10.1063/1.4916924}
}

@article{bassett2017network,
  title={Network neuroscience},
  author={Bassett, Danielle S and Sporns, Olaf},
  journal={Nature Neuroscience},
  volume={20},
  number={3},
  pages={353--364},
  year={2017},
  publisher={Nature Publishing Group US New York},
  doi={10.1038/nn.4502},
  URL={https://www.nature.com/articles/nn.4502}
}

@article{tandon2023multilayer,
  title={Multilayer network analysis to study complex inter-subsystem interactions in a turbulent thermoacoustic system},
  author={Tandon, Shruti and Sujith, Raman I},
  journal={Journal of Fluid Mechanics},
  volume={966},
  pages={A9},
  year={2023},
  publisher={Cambridge University Press},
  doi={10.1017/jfm.2023.338},
  url={https://doi.org/10.1017/jfm.2023.338}
}

@article{shri2022complex,
  title={A complex network framework for studying particle-laden flows},
  author={Shri Vignesh, K and Tandon, Shruti and Kasthuri, Praveen and Sujith, RI},
  journal={Physics of Fluids},
  volume={34},
  number={7},
  pages={073321},
  year={2022},
  publisher={AIP Publishing},
  doi={10.1063/5.0098917},
  url={https://doi.org/10.1063/5.0098917}
}

@inproceedings{yang2013overlapping,
  title={Overlapping community detection at scale: a nonnegative matrix factorization approach},
  author={Yang, Jaewon and Leskovec, Jure},
  booktitle={Proceedings of the sixth ACM international conference on Web search and data mining},
  pages={587--596},
  year={2013},
  doi={10.1145/2433396.2433471},
  url={https://doi.org/10.1145/2433396.2433471}
}

@inbook{kelley2012defining,
  title={Defining and {D}iscovering {C}ommunities in {S}ocial {N}etworks},
  author={Kelley, Stephen and Goldberg, Mark and Magdon-Ismail, Malik and Mertsalov, Konstantin and Wallace, Al},
  booktitle={Handbook of Optimization in Complex Networks: Theory and Applications},
  pages={139--168},
  year={2012},
  publisher={Springer},
  doi={10.1007/978-1-4614-0754-6_6},
  url={https://doi.org/10.1007/978-1-4614-0754-6_6}
}

@article{lancichinetti2009community,
  title={Community detection algorithms: {A} comparative analysis},
  author={Lancichinetti, Andrea and Fortunato, Santo},
  journal={Physical Review E},
  volume={80},
  number={5},
  pages={056117},
  year={2009},
  publisher={APS},
  doi={10.1103/PhysRevE.80.056117},
  url={https://doi.org/10.1103/PhysRevE.80.056117}
}

@article{newman2004finding,
  title={Finding and evaluating community structure in networks},
  author={Newman, Mark EJ and Girvan, Michelle},
  journal={Physical Review E},
  volume={69},
  number={2},
  pages={026113},
  year={2004},
  publisher={APS},
  doi={10.1103/PhysRevE.69.026113},
  url={https://doi.org/10.1103/PhysRevE.69.026113}
}

@article{newman2004fast,
  title={Fast algorithm for detecting community structure in networks},
  author={Newman, Mark EJ},
  journal={Physical Review E},
  volume={69},
  number={6},
  pages={066133},
  year={2004},
  publisher={APS},
  doi={10.1103/PhysRevE.69.066133},
  url={https://doi.org/10.1103/PhysRevE.69.066133}
}

@article{blondel2008fast,
  title={Fast unfolding of communities in large networks},
  author={Blondel, Vincent D and Guillaume, Jean-Loup and Lambiotte, Renaud and Lefebvre, Etienne},
  journal={Journal of Statistical Mechanics: Theory and Experiment},
  volume={2008},
  number={10},
  pages={P10008},
  year={2008},
  publisher={IOP Publishing},
  doi={10.1088/1742-5468/2008/10/P10008},
  url={https://doi.org/10.1088/1742-5468/2008/10/P10008}
}

@article{prekopa1957poisson,
  title={On {P}oisson and composed {P}oisson stochastic set functions},
  author={Pr{\'e}kopa, Andr{\'a}s},
  journal={Studia Math},
  volume={16},
  pages={142--155},
  year={1957},
  doi={10.4064/sm-16-2-142-155},
  url={https://rutcor.rutgers.edu/~prekopa/poisson.pdf}
}

@article{fischer1993markov,
  title={The {M}arkov-modulated {P}oisson process ({MMPP}) cookbook},
  author={Fischer, Wolfgang and Meier-Hellstern, Kathleen},
  journal={Performance Evaluation},
  volume={18},
  number={2},
  pages={149--171},
  year={1993},
  publisher={Elsevier},
  doi={10.1016/0166-5316(93)90035-S},
  url={https://doi.org/10.1016/0166-5316(93)90035-S}
}

@article{carrara2015heart,
  title={Heart rate dynamics distinguish among atrial fibrillation, normal sinus rhythm and sinus rhythm with frequent ectopy},
  author={Carrara, Marta and Carozzi, Luca and Moss, Travis J and De Pasquale, Marco and Cerutti, Sergio and Ferrario, Manuela and Lake, Douglas E and Moorman, J Randall},
  journal={Physiological measurement},
  volume={36},
  number={9},
  pages={1873},
  year={2015},
  publisher={IOP Publishing},
  doi={10.1088/0967-3334/36/9/1873},
  url={https://doi.org/10.1088/0967-3334/36/9/1873}
}

@article{tateno2001automatic,
  title={Automatic detection of atrial fibrillation using the coefficient of variation and density histograms of {RR} and $\Delta${RR} intervals},
  author={Tateno, K and Glass, L},
  journal={Medical and Biological Engineering and Computing},
  volume={39},
  pages={664--671},
  year={2001},
  publisher={Springer},
  doi={10.1007/BF02345439},
  url={https://doi.org/10.1007/BF02345439}
}

@article{goldberger2000physiobank,
  title={Physio{B}ank, {P}hysio{T}oolkit, and {P}hysioNet: components of a new research resource for complex physiologic signals},
  author={Goldberger, Ary L and Amaral, Luis AN and Glass, Leon and Hausdorff, Jeffrey M and Ivanov, Plamen Ch and Mark, Roger G and Mietus, Joseph E and Moody, George B and Peng, Chung-Kang and Stanley, H Eugene},
  journal={Circulation},
  volume={101},
  number={23},
  pages={e215--e220},
  year={2000},
  publisher={Am Heart Assoc},
  doi={10.1161/01.cir.101.23.e215},
  url={https://doi.org/10.1161/01.cir.101.23.e215}
}

@article{moody2001impact,
  title={The impact of the {MIT-BIH} arrhythmia database},
  author={Moody, George B and Mark, Roger G},
  journal={IEEE Engineering in Medicine and Biology magazine},
  volume={20},
  number={3},
  pages={45--50},
  year={2001},
  publisher={IEEE},
  doi={10.1109/51.932724},
  url={https://doi.org/10.1109/51.932724}
}

@article{chaumat2001droplet,
  title={Droplet {S}pectra {B}roadening in {C}umulus {C}louds. {P}art {II}: {M}icroscale {D}roplet {C}oncentration {H}eterogeneities},
  author={Chaumat, Laure and Brenguier, Jean-Louis},
  journal={Journal of the Atmospheric Sciences},
  volume={58},
  number={6},
  pages={642--654},
  year={2001},
  publisher={American Meteorological Society},
  doi={10.1175/1520-0469(2001)058<0642:DSBICC>2.0.CO;2},
  url={https://doi.org/10.1175/1520-0469(2001)058<0642:DSBICC>2.0.CO;2}
}

@article{marshak2005small,
  title={Small-{S}cale {D}rop-{S}ize {V}ariability: {E}mpirical {M}odels for {D}rop-{S}ize-{D}ependent {C}lustering in {C}louds},
  author={Marshak, Alexander and Knyazikhin, Yuri and Larsen, Michael L and Wiscombe, Warren J},
  journal={Journal of the Atmospheric Sciences},
  volume={62},
  number={2},
  pages={551--558},
  year={2005},
  publisher={American Meteorological Society},
  doi={10.1175/JAS-3371.1},
  url={https://doi.org/10.1175/JAS-3371.1}
}

@article{kostinski2001scale,
  title={Scale-dependent droplet clustering in turbulent clouds},
  author={Kostinski, Alexander B and Shaw, Raymond A},
  journal={Journal of Fluid Mechanics},
  volume={434},
  pages={389--398},
  year={2001},
  publisher={Cambridge University Press},
  doi={10.1017/S0022112001004001},
  url={https://doi.org/10.1017/S0022112001004001}
}

@article{lehmann2007evidence,
  title={Evidence for inertial droplet clustering in weakly turbulent clouds},
  author={Lehmann, Katrin and Siebert, Holger and Wendisch, Manfred and Shaw, Raymond A},
  journal={Tellus B: Chemical and Physical Meteorology},
  volume={59},
  number={1},
  pages={57--65},
  year={2007},
  publisher={Taylor \& Francis},
  doi={10.1111/j.1600-0889.2006.00229.x},
  url={https://doi.org/10.1111/j.1600-0889.2006.00229.x}
}

@article{larsen2014recovery,
  title={On the {R}ecovery of 3{D} {S}patial {S}tatistics of {P}articles from 1{D} {M}easurements: {I}mplications for {A}irborne {I}nstruments},
  author={Larsen, Michael L and Briner, Clarissa A and Boehner, Philip},
  journal={Journal of Atmospheric and Oceanic Technology},
  volume={31},
  number={10},
  pages={2078--2087},
  year={2014},
  publisher={American Meteorological Society},
  doi={10.1175/JTECH-D-14-00004.1},
  url={https://doi.org/10.1175/JTECH-D-14-00004.1}
}

@article{uhlig1998holographic,
  title={Holographic in-situ measurements of the spatial droplet distribution in stratiform clouds},
  author={Uhlig, Eva-Maria and Borrmann, Stephan and Jaenicke, Ruprecht},
  journal={Tellus B: Chemical and Physical Meteorology,},
  volume={50},
  number={4},
  pages={377--387},
  year={1998},
  publisher={Wiley Online Library},
  doi={10.3402/tellusb.v50i4.16210},
  url={https://doi.org/10.3402/tellusb.v50i4.16210}
}

@article{jameson2000fluctuation,
  title={Fluctuation {P}roperties of {P}recipitation. {P}art {VI}: {O}bservations of {H}yperfine {C}lustering and {D}rop {S}ize {D}istribution {S}tructures in {T}hree-{D}imensional {R}ain},
  author={Jameson, AR and Kostinski, AB},
  journal={Journal of the Atmospheric Sciences},
  volume={57},
  number={3},
  pages={373--388},
  year={2000},
  publisher={American Meteorological Society},
  doi={10.1175/1520-0469(2000)057<0373:FPOPPV>2.0.CO;2},
  url={https://doi.org/10.1175/1520-0469(2000)057<0373:FPOPPV>2.0.CO;2}
}

@article{fugal2009cloud,
  title={Cloud particle size distributions measured with an airborne digital in-line holographic instrument},
  author={Fugal, JP and Shaw, RA},
  journal={Atmospheric Measurement Techniques},
  volume={2},
  number={1},
  pages={259--271},
  year={2009},
  publisher={Copernicus GmbH},
  doi={10.5194/amt-2-259-2009},
  url={https://doi.org/10.5194/amt-2-259-2009}
}

@article{larsen2018fine,
  title={Fine-scale droplet clustering in atmospheric clouds: 3{D} radial distribution function from airborne digital holography},
  author={Larsen, Michael L and Shaw, Raymond A and Kostinski, Alexander B and Glienke, Susanne},
  journal={Physical Review Letters},
  volume={121},
  number={20},
  pages={204501},
  year={2018},
  publisher={APS},
  doi={10.1103/PhysRevLett.121.204501},
  url={https://doi.org/10.1103/PhysRevLett.121.204501}
}

@article{fruchterman1991graph,
  title={Graph drawing by force-directed placement},
  author={Fruchterman, Thomas MJ and Reingold, Edward M},
  journal={Software: Practice and Experience},
  volume={21},
  number={11},
  pages={1129--1164},
  year={1991},
  publisher={Wiley Online Library},
  doi={10.1002/spe.4380211102},
  url={https://doi.org/10.1002/spe.4380211102}
}

@inproceedings{bastian2009gephi,
  title={Gephi: an open source software for exploring and manipulating networks},
  author={Bastian, Mathieu and Heymann, Sebastien and Jacomy, Mathieu},
  booktitle={Proceedings of the international AAAI conference on web and social media},
  volume={3},
  pages={361--362},
  year={2009},
  doi={10.1609/icwsm.v3i1.13937},
  url={https://doi.org/10.1609/icwsm.v3i1.13937 }
}

@book{cinlar1975introduction,
  title={Introduction to stochastic processes},
  author={Cinlar, Erhan},
  year={2013},
  publisher={Courier Corporation},
  doi={10.1137/1019090},
  url={https://doi.org/10.1137/1019090}
}

@article{ryden1996algorithm,
  title={An {EM} algorithm for estimation in {M}arkov-modulated {P}oisson processes},
  author={Ryd{\'e}n, Tobias},
  journal={Computational Statistics \& Data Analysis},
  volume={21},
  number={4},
  pages={431--447},
  year={1996},
  publisher={Elsevier},
  doi={10.1016/0167-9473(95)00025-9},
  url={https://doi.org/10.1016/0167-9473(95)00025-9}
}

@article{monchaux2010preferential,
  title={Preferential concentration of heavy particles: a Vorono{\"\i} analysis},
  author={Monchaux, Romain and Bourgoin, Micka{\"e}l and Cartellier, Alain},
  journal={Physics of Fluids},
  volume={22},
  number={10},
  year={2010},
  publisher={AIP Publishing},
  doi={10.1063/1.3489987},
  url={https://doi.org/10.1063/1.3489987}
}

@book{benes2017general,
  title={General stochastic processes in the theory of queues},
  author={Benes, Vaclav E},
  year={2017},
  publisher={Courier Dover Publications},
  url={https://books.google.co.in/books/about/General_Stochastic_Processes_in_the_Theo.html?id=LTc4AAAAMAAJ&redir_esc=y}
}

@book{ibe2013markov,
  title={Markov processes for stochastic modeling},
  author={Ibe, Oliver},
  year={2013},
  publisher={Newnes},
  doi={10.1016/C2012-0-06106-6},
  url={https://doi.org/10.1016/C2012-0-06106-6}
}

@article{sumbekova2017preferential,
  title={Preferential concentration of inertial sub-Kolmogorov particles: the roles of mass loading of particles, Stokes numbers, and Reynolds numbers},
  author={Sumbekova, Sholpan and Cartellier, Alain and Aliseda, Alberto and Bourgoin, Mickael},
  journal={Physical Review Fluids},
  volume={2},
  number={2},
  pages={024302},
  year={2017},
  publisher={APS},
  doi={10.1103/PhysRevFluids.2.024302},
  url={https://doi.org/10.1103/PhysRevFluids.2.024302}
}

@article{obligado2014preferential,
  title={Preferential concentration of heavy particles in turbulence},
  author={Obligado, Martin and Teitelbaum, Tomas and Cartellier, Alain and Mininni, Pablo and Bourgoin, Mickael},
  journal={Journal of Turbulence},
  volume={15},
  number={5},
  pages={293--310},
  year={2014},
  publisher={Taylor \& Francis},
  doi={10.1080/14685248.2014.897710},
  url={https://doi.org/10.1080/14685248.2014.897710}
}

@article{ferenc2007size,
  title={On the size distribution of {P}oisson {V}oronoi cells},
  author={Ferenc, J{\'a}rai Szab{\'o} and N{\'e}da, Zolt{\'a}n},
  journal={Physica A: Statistical Mechanics and its Applications},
  volume={385},
  number={2},
  pages={518--526},
  year={2007},
  publisher={Elsevier},
  doi={10.1016/j.physa.2007.07.063},
  url={https://doi.org/10.1016/j.physa.2007.07.063}
}

@book{albrecht2013laser,
  title={Laser Doppler and phase Doppler measurement techniques},
  author={Albrecht, H-E and Damaschke, Nils and Borys, Michael and Tropea, Cameron},
  year={2013},
  publisher={Springer Science \& Business Media},
  doi={10.1007/978-3-662-05165-8},
  url={https://doi.org/10.1007/978-3-662-05165-8}
}

@article{saikranthi2013identification,
  title={Identification and validation of homogeneous rainfall zones in {I}ndia using correlation analysis},
  author={Saikranthi, K and Rao, T Narayana and Rajeevan, M and Bhaskara Rao, S Vijaya},
  journal={Journal of Hydrometeorology},
  volume={14},
  number={1},
  pages={304--317},
  year={2013},
  doi={10.1175/JHM-D-12-071.1},
  url={https://journals.ametsoc.org/view/journals/hydr/14/1/jhm-d-12-071_1.xml}
}

@article{ShriVigneshTurb,
  title={Turbulence is ineffective in causing raindrop growth in polluted clouds},
  author={ Shri Vignesh, K and Sukdeo, Ambedkar Sanket and Sruthibhai, P V and Singh, Aishwarya and Sahu, Srikrishna and  Chaudhari, Swetaprovo and Patra, Amit K and Rao, T Narayana and Govindarajan, Rama and Gunthe, Sachin S and Sujith, R I},
  journal={arXiv:2601.00637},
  year={2026},
  doi = {10.48550/arXiv.2601.00637},
  url={https://doi.org/10.48550/arXiv.2601.00637}
}

@article{abdi2010coefficient,
  title={Coefficient of variation},
  author={Abdi, Herv{\'e}},
  journal={Encyclopedia of research design},
  volume={1},
  number={5},
  pages={169--171},
  year={2010},
  url={https://personal.utdallas.edu/~herve/abdi-cv2010-pretty.pdf}
}

@article{SU20151,
title = {A method for discovering clusters of e-commerce interest patterns using click-stream data},
author = {Qiang Su and Lu Chen},
journal = {Electronic Commerce Research and Applications},
volume = {14},
number = {1},
pages = {1--13},
year = {2015},
doi = {https://doi.org/10.1016/j.elerap.2014.10.002},
url = {https://www.sciencedirect.com/science/article/pii/S1567422314000726}
}

@article{singh2021clustering,
  author  = {Singh, Harpreet and Kaur, Parminder},
  title   = {An Effective Clustering-Based Web Page Recommendation Framework for E-Commerce Websites},
  journal = {SN Computer Science},
  volume  = {2},
  number  = {4},
  pages   = {339},
  year    = {2021},
  publisher = {Springer},
  doi     = {10.1007/s42979-021-00697-3},
  url = {https://link.springer.com/content/pdf/10.1007/s42979-021-00736-z.pdf}
}

@article{wuebben2016getting,
  author  = {Wuebben, Daniel},
  title   = {Getting Likes, Going Viral, and the Intersections Between Popularity Metrics and Digital Composition},
  journal = {Computers and Composition},
  volume  = {42},
  pages   = {1--12},
  year    = {2016},
  publisher = {Elsevier},
  doi     = {10.1016/j.compcom.2016.09.002},
  url = {https://www.sciencedirect.com/science/article/pii/S8755461515300098}
}

@book{Kantz2004,
  author = {Kantz, Holger and Schreiber, Thomas},
  title = {Nonlinear Time Series Analysis},
  publisher = {Cambridge University Press},
  year = {2004},
  doi = {10.1017/CBO9780511755798},
  url = {https://www.cambridge.org/core/books/nonlinear-time-series-analysis/519783E4E8A2C3DCD4641E42765309C7}
}

@article{dakos2024tipping,
  title={Tipping point detection and early warnings in climate, ecological, and human systems},
  author={Dakos, Vasilis and Boulton, Chris A and Buxton, Joshua E and Abrams, Jesse F and Arellano-Nava, Beatriz and Armstrong McKay, David I and Bathiany, Sebastian and Blaschke, Lana and Boers, Niklas and Dylewsky, Daniel and others},
  journal={Earth System Dynamics},
  volume={15},
  number={4},
  pages={1117--1135},
  year={2024},
  publisher={Copernicus GmbH},
  doi= {10.5194/esd-15-1117-2024},
  url = {https://esd.copernicus.org/articles/15/1117/2024/esd-15-1117-2024.html}
}

@article{yang2022critical,
  title={Critical transitions in the hydrological system: early-warning signals and network analysis},
  author={Yang, Xueli and Wang, Zhi-Hua and Wang, Chenghao},
  journal={Hydrology and Earth System Sciences},
  volume={26},
  number={7},
  pages={1845--1856},
  year={2022},
  publisher={Copernicus GmbH},
  doi = {10.5194/hess-26-1845-2022},
  url = {https://hess.copernicus.org/articles/26/1845/2022/}
}

@article{banerjee2024early,
  title={Early warnings of tipping in a non-autonomous turbulent reactive flow system: Efficacy, reliability, and warning times},
  author={Banerjee, Ankan and Pavithran, Induja and Sujith, RI},
  journal={Chaos: An Interdisciplinary Journal of Nonlinear Science},
  volume={34},
  number={1},
  pages={013113},
  year={2024},
  publisher={AIP Publishing},
  doi = {10.1063/5.0160918},
  url = {https://pubs.aip.org/aip/cha/article/34/1/013113/2932959}
}

@article{radhakrishnan2025early,
  title={Early warnings are too late when parameters change rapidly},
  author={Radhakrishnan, Rohit and Pavithran, Induja and Livina, Valerie and Kurths, J{\"u}rgen and Sujith, RI},
  journal={Scientific Reports},
  volume={15},
  number={1},
  pages={20256},
  year={2025},
  publisher={Nature Publishing Group UK London},
  doi = {10.1038/s41598-025-06525-5},
  url = {https://www.nature.com/articles/s41598-025-06525-5.pdf}
}

@article{pace2017reversal,
  title={Reversal of a cyanobacterial bloom in response to early warnings},
  author={Pace, Michael L and Batt, Ryan D and Buelo, Cal D and Carpenter, Stephen R and Cole, Jonathan J and Kurtzweil, Jason T and Wilkinson, Grace M},
  journal={Proceedings of the National Academy of Sciences},
  volume={114},
  number={2},
  pages={352--357},
  year={2017},
  publisher={National Academy of Sciences},
  doi = {10.1073/pnas.1612424114},
  url = {https://www.pnas.org/doi/pdf/10.1073/pnas.1612424114}
}

\end{document}